\documentclass[lettersize,journal]{IEEEtran}
\usepackage{amsmath,amsfonts}
\usepackage{algorithmic}
\usepackage{algorithm}
\usepackage{array}
\usepackage[caption=false,font=normalsize,labelfont=sf,textfont=sf]{subfig}
\usepackage{textcomp}
\usepackage{stfloats}
\usepackage{url}
\usepackage{verbatim}
\usepackage{graphicx}
\usepackage{cite}
\usepackage{subfig}
\usepackage{orcidlink}
\usepackage{gensymb}
\definecolor{cgcol}{rgb}{0,0,0}

\begin{document}

\title{Distributed Stochastic Optimization of a Neural Representation Network for Time-Space Tomography Reconstruction}

\author{K.~Aditya~Mohan\textsuperscript{\orcidlink{0000-0002-0921-6559}},~Massimiliano~Ferrucci\textsuperscript{\orcidlink{0000-0002-8811-8681}},~Chuck~Divin\textsuperscript{\orcidlink{0000-0001-6067-2288}},~Garrett~A.~Stevenson\textsuperscript{\orcidlink{0000-0001-7085-8334}},~Hyojin~Kim\textsuperscript{\orcidlink{0000-0001-7032-0999}}
% <-this % stops a space
\thanks{K. A. Mohan, M. Ferrucci, C. Divin, G. A. Stevenson, and H. Kim are affiliated with 
Lawrence Livermore National Laboratory, Livermore, CA. This paper has supplementary downloadable material available at http://ieeexplore.ieee.org., provided by the author.}}% <-this % stops a space
%\thanks{Manuscript received April 19, 2021; revised August 16, 2021.}}



\maketitle

\begin{abstract}
4D time-space reconstruction of 
dynamic events or deforming objects using X-ray computed tomography (CT) 
is an \textcolor{cgcol}{important inverse problem in non-destructive evaluation.} \textcolor{cgcol}{Conventional back-projection based reconstruction methods
assume that the object remains static for the duration of several 
tens or hundreds of X-ray projection measurement images 
(reconstruction of consecutive limited-angle CT scans).}
However, this is an unrealistic assumption for many 
in-situ experiments that causes spurious artifacts
and inaccurate morphological reconstructions of the object.
To solve this problem, we propose to perform a 
4D time-space reconstruction 
using a distributed implicit neural representation (DINR) network 
that is trained using a novel distributed 
stochastic training algorithm. 
Our DINR network learns to reconstruct the object at its output by 
iterative optimization of its network parameters such that 
the measured projection images best match the 
output of the CT forward measurement model. 
We use a forward measurement
model that is a function of the DINR outputs at a sparsely 
sampled set of continuous valued \textcolor{cgcol}{4D} object coordinates.
Unlike previous neural representation
architectures that forward and back propagate through 
dense voxel grids that sample the object's entire time-space
coordinates, we only propagate through the DINR
at a small subset of object coordinates in each iteration 
resulting in an order-of-magnitude reduction 
in memory and compute for training. 
DINR leverages distributed computation 
across several compute nodes and GPUs to produce 
high-fidelity 4D time-space reconstructions.
We use both simulated parallel-beam 
and experimental cone-beam X-ray CT datasets
to demonstrate the superior performance of our approach.
\end{abstract}


%\section{Introduction}\label{sec:intro}

%(Placeholder text for this section) 4D Time-Space Computed Tomography (4DCT) is an emerging imaging modality for non-destructive characterization of dynamic 3-dimensional (3D) objects in scientific imaging. 4DCT has been used for 3D reconstruction of the temporal dynamics in scientific imaging experiments such as dendritic solidification \cite{}, battery failure degration \cite{}, etc. 
%In X-ray CT, an object is rotated along an axis and measurements are acquired at several view angles of the rotating object. The popular approaches to CT reconstruction rely on an assumption of static object for the duration of several projections over a wide angular range of rotation (say $180^0$). Thus, the time axis is split into several contiguous non-overlapping frames and one 3D volume is reconstructed using the projection data within each frame.
% In reality, however, the object changes continuously from view-to-view due to the temporal evolution of the dynamic object. Hence, the current approaches for 4DCT are applicable only if the temporal evolution is sufficiently slow to make an assumption of static object within the duration of a single frame. This is a major limitation of existing 4DCT approaches since it imposes a lower limit on the achievable temporal resolution that is equal to the frame duration.

\section{Introduction}
X-ray computed tomography (CT) is a widely used imaging modality for 
non-destructive characterization in industrial and scientific imaging, 
clinical diagnosis in medical imaging, 
and border security at airports. 
It is useful to produce 3D object reconstructions from 
X-ray projection measurement images at several view angles 
around the object. 
%Tomographic 3D reconstruction of an object from its projection 
%images is a linear inverse problem with numerous solutions 
%that include analytical back-projection methods ~\cite{kak1998},
%iterative algorithms ~\cite{mohan2015timbir}, and deep learning approaches ~\cite{wang2020}. 
Unlike conventional CT, 4D time-space CT (4DCT) is an 
emerging imaging modality that is useful 
for reconstruction of dynamic scenes or deforming objects.
However, 4DCT poses a challenging and ill-posed inverse problem 
since the X-ray projection images are snapshots of a 
continuously time varying 3D scene.
4DCT has been used for 
studying in-situ object deformation under mechanical and thermal loading,
dendritic solidification ~\cite{gibbs2015three, mohan2015timbir}, 
battery failure degradation ~\cite{Kok_2023,ziesche20204d}, 
and periodic motion of the breathing phases in clinical diagnosis 
~\cite{taubmann2017spatio, schwemmer2013opening, schwemmer2013residual, rohkohl2008c, keall2004acquiring, pan20044d}. 
Several 4DCT methods have also been proposed to reconstruct 
non-periodic deforming scenes using 
motion field parameters ~\cite{zang2019warp, zang2018space}. 
\textcolor{cgcol}{While our proposed approach is generally applicable, 
our goal in this paper is 4DCT reconstruction of fast dynamics in objects 
that are compressed by more than half the detector's 
2D field-of-view from only few hundreds of view angles acquired 
over one to four number of $180^\circ$ rotations.
Furthermore, our experimental data reconstructions occupy several
terabytes of storage space due to the reconstruction
of one volume per projection view during inference.}

\textcolor{cgcol}{
In conventional X-ray CT, the 3D reconstruction approaches
based on analytical back-projections 
rely on the assumption of a static scene during 
the entire projection data acquisition \cite{Feldkamp84_FDKrecon, kak1998}.} 
In 4DCT, however, the scene is expected to change 
continuously over time and between subsequent views.
Such a temporal change may be attributed to several factors
including object motion, system instabilities, 
and dynamic physical processes such as 
thermal or mechanical loading. 
The \textcolor{cgcol}{conventional approach to 4DCT 
is to group the projection images into several time frames such 
that each frame consists of consecutive projection images 
over a predefined angular range.
This angular range is typically $180^{\circ}$ for parallel-beam CT
and range from $180^{\circ}$ to $360^{\circ}$ for cone-beam CT \cite{Feldkamp84_FDKrecon, kak1998}.}
For each time frame, a tomographic reconstruction algorithm is used to reconstruct a 3D volume from its corresponding projections.

\textcolor{cgcol}{We avoid limited angle artifacts in the
reconstructed 3D volumes \cite{mohan2015timbir} 
by ensuring that each frame contains} projections acquired over 
a minimum rotation of $180^{\circ}$.
Note that this requirement is \textcolor{cgcol}{applicable} to both 
static and dynamic CT.
The projections are grouped such that changes in the scene 
within each frame are small enough that the scene can 
be considered static for the duration of the frame. 
This condition limits the achievable temporal resolution 
to the duration of each frame of projections 
and is therefore not suitable for rapidly changing scenes. 
Thus, conventional reconstruction approaches for 4DCT 
produce inaccurate reconstructions that are
deteriorated by spurious artifacts and blurry features 
when the object changes rapidly within a frame.

\textcolor{cgcol}{Two classes of iterative algorithms have been proposed 
to address the limitation of necessitating $180^{\circ}$ rotation for each frame.
The first class of algorithms use prior models to enforce
sparsity of reconstructions along the space and time dimensions.
These prior models may penalize differences between the values
of neighboring voxels \cite{mohan2015timbir} or penalize differences
between non-local patches of voxels \cite{Kazantsev_2015_InvProblems}.
PICCS \cite{Chen_2008_MedPhy} constrains the dynamics to small
deviations around a prior static image that is reconstructed from a combined dataset of all the dynamic CT scans.
%These prior models essentially assume a multi-variate probability density
%function (such as Gibbs distribution) for the distribution of voxel values.
The second class of algorithms use basis functions or motion models
to explicitly model the time dependence of voxel values. 
These models include piecewise constant (PWC) functions \cite{Eyndhoven_2015_TIP},
piecewise linear (PWL) functions \cite{Boigne_2022_TCI}, and
Fourier basis functions \cite{Nikitin_2019_TCI}.
These second class of algorithms can provide substantial benefits if the explicit analytical formulas for the motion models are a high-fidelity representation for the unknown dynamics.
\cite{Clement_2018_Materials} proposes an adaptive algorithmic approach 
to estimate a time/space displacement field.
%during reconstruction 
%for limited complexity and amplitude of displacements. 
The use of such parameterized motion models can be beneficial or detrimental depending
on whether they accurately or inaccurately model the object's dynamics respectively.}

Implicit neural representation (INR) networks,
also called neural radiance fields (NeRF), have demonstrated
remarkable potential in solving several inverse imaging problems. 
%such as view synthesis, object reconstruction, 
%object pose estimation, and other physics simulations. 
They have been successfully applied in several applications
including view synthesis  ~\cite{Mildenhall20eccv_nerf, riegler2020free, dupont2020equivariant, zhang2021neural, wang2022r2l}, 
texture completion ~\cite{chibane2020implicit}, 
deformable scene estimation ~\cite{park2020deformable, pumarola2020d}, 
and 3D reconstruction ~\cite{yariv2021volume, wang2021neus, meng_2023_neat}. 
INR networks behave as function approximators by representing
an object's physical or material properties as a differentiable 
function of the 3D spatial or temporal coordinates.
\textcolor{cgcol}{A popular choice for INR} is to use fully connected layers 
(multi-layer perceptron) whose parameters are trained to learn
a \textcolor{cgcol}{suitable} mapping from coordinates to object properties.
\textcolor{cgcol}{Such a mapping} is learned by minimizing a loss function whose purpose
is to minimize the discrepancy between the measured data
and a forward model, which is a differentiable
mathematical model for the sensor or detector data as a function
of the properties of the object.

In recent years, several INR-based approaches \textcolor{cgcol}{\cite{Reed2021, shao2024dynamic, Lee_2024_INR, Shen_2024_NERP}} have been 
successfully applied to various CT reconstruction problems.
\textcolor{cgcol}{For comparisons, we choose an INR 
coupled with parametric motion fields \cite{Reed2021}}. 
This approach is useful for reconstruction of deforming objects
or periodic motions by optimizing both an object template and motion field parameters
used for representing the temporal dynamics by appropriate warping of the template.
While this approach significantly outperformed other SOTA approaches, 
its implementation is prohibitively expensive in terms of the computational
and memory resource requirements for the Graphics Processing Units (GPUs).
This resource limitation is primarily due to forward and back propagation
through discrete voxel\footnote{Voxel is an abbreviation of ``volumetric picture element'' and is the 3D analog of a 2D pixel.} 
grids that fully instantiate the object properties across space and time.
Hence, this approach is infeasible for real-world use-cases
where it may not be possible to even store a 
single 3D voxelated volume in a GPU's memory.
For 4DCT, we may require storage of several hundreds of voxelated 
volumes that represent the object's dynamics over time. 
We estimate that thousands to tens of thousands of GPU resources may be required 
for 4DCT reconstructions of realistic experimental data that may span
gigabytes to terabytes in size.

In this paper, we present a novel reconstruction approach 
called Distributed Implicit Neural Representation (DINR) 
that uses a new distributed training algorithm for 
optimizing a continuous space-time representation of the object. 
DINR can scale to \textcolor{cgcol}{reconstruct large data sizes of more than a billion projection pixels}
since it only uses \textcolor{cgcol}{few thousands} of randomly sampled projection pixel 
values for optimization of the network parameters in any training iteration.
For each randomly chosen projection pixel, DINR
requires forward and back propagation through the network
at only those 4D object coordinates
that contribute to the sampled projection pixels.
Thus, the memory and computational requirements per GPU
 is drastically lower than previous INRs that
instantiate large sized voxel grids during training.
DINR learns to reconstruct the object in 4D at high resolutions 
using multiple GPUs that are distributed over several nodes.
The number of sampled projection pixels can be adjusted to fit 
the maximum combined memory capacity of the available GPUs. 
Our approach can be run either on single GPU machines, 
multi-GPU workstations, or multi-node GPU clusters 
in high-performance computing (HPC) architectures.

One of our key contributions is the use of a
continuous representation for the forward model that is 
purposefully designed for accuracy and ease of distributed training. 
Training using ray tracing \cite{Bello_ProNERF_2024, Barron_MIPNERF_2021, Otonari_NUNERF_2022}, 
i.e., one ray for each sensor pixel, 
is the state-of-the-art in NERF for enabling distributed stochastic optimization.
In CT, however, INRs {\cite{Reed2021, shao2024dynamic, Lee_2024_INR, Shen_2024_NERP}} have not used ray tracing and instead opted for discrete forward models (also called system matrices)
that have evolved from fast but less accurate methods 
such as Siddon’s method \cite{siddon1985fast}, 
to more sophisticated methods such as separable footprints \cite{Long_2010_SF}
that model the impact of discrete voxel grids on the projections. 
\textcolor{cgcol}{However, these models require the instantiation of multiple voxelated\footnote{Voxels are cuboids with a non-zero volume and projectors for voxels approximate the average line-integral over all rays propagating through this voxel.} volumes, each representing a specific time step. As a result, when the number of time steps is large, the sequence of volumes to be instantiated grows significantly.}
Hence, the traditional approach to CT reconstruction using INR 
\cite{Reed2021, shao2024dynamic, Lee_2024_INR, Shen_2024_NERP}
is memory limited by the space necessary for storing
the volumes.
In this paper, we propose to use a small random subset of projection
pixels in each iteration that only require sampling 
of 3D points in the object that lie within the pyramidal volume 
whose apex is the X-ray source and the base
is the plane of a single sensor pixel.
This approach is much more accurate for CT applications
than the traditional NERF approach of propagating a single ray 
for each sensor pixel. Lastly, we demonstrate our approach using real experimental 
4DCT datasets of two additively manufactured parts. 
We released these datasets publicly under an open source license \cite{DebenDataset}. 

\begin{figure*}[!thb]
\begin{center}
\includegraphics[width=6in]{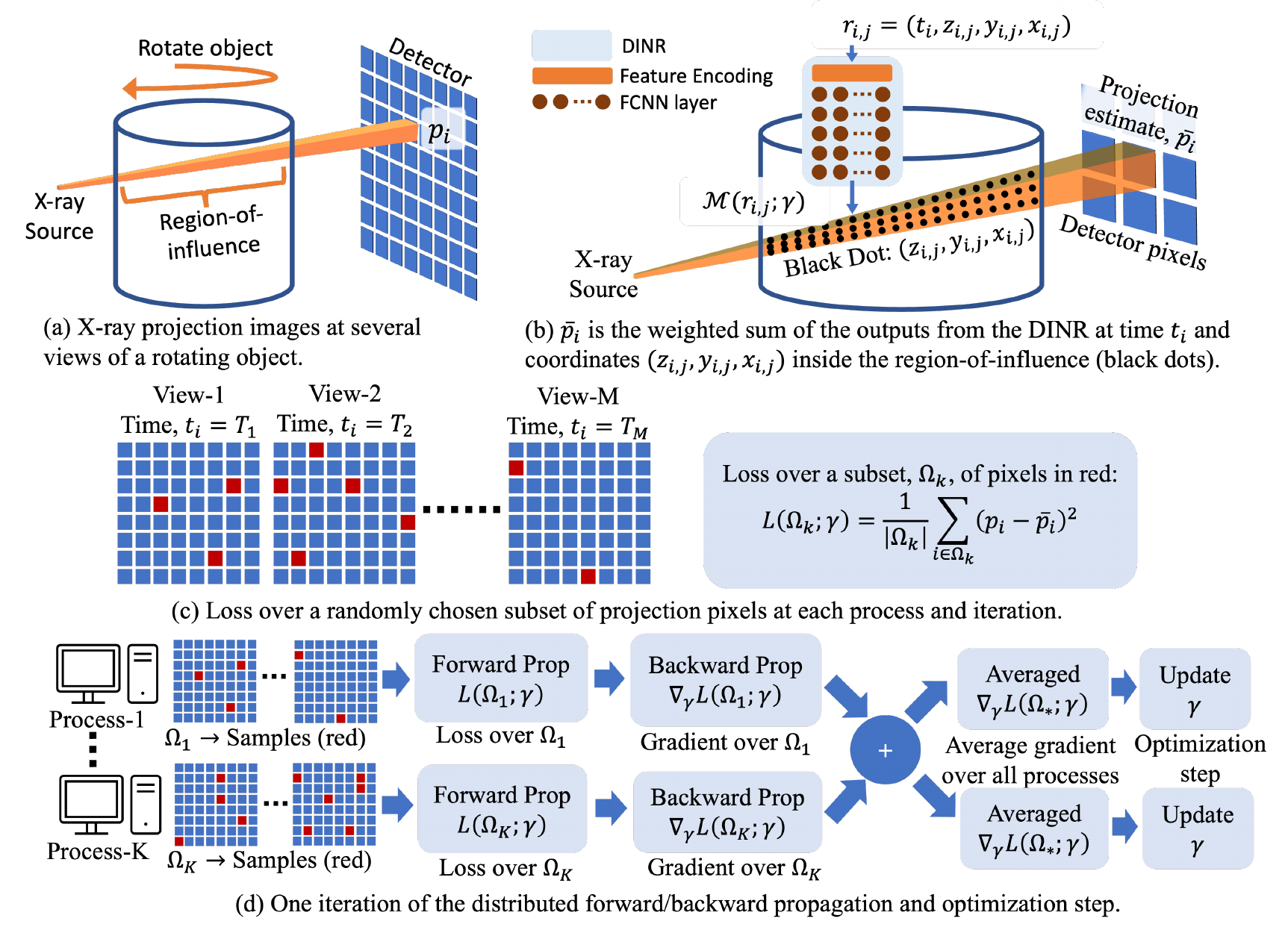}
\end{center}
\caption{\label{fig:mainfig}
Schematic of our distributed implicit neural representation 
(DINR) approach to 4DCT reconstruction.
The projection estimate $\bar{p}_i$ is a function of the 
linear attenuation coefficient (LAC) at coordinates
$r_{i,j}$ inside the orange colored 
pyramidal volume shown in (a, b).
The output of the DINR network, 
$\mathcal{M}(r_{i,j}; \gamma)$, gives the LAC 
at coordinate $r_{i,j}$.
(c) shows the loss function that is local
to each process and computed over a small subset 
$\Omega_k$ of projection indices.
For the projection image at the $m^{th}$ view,
the time $t_i$ for each projection pixel 
is the same value of $T_m$ since the time instant
for all pixels in an image is the same.
(d) is our distributed approach
to training of the DINR network.
}
\end{figure*}

\section{Our Approach}
A schematic of our experimental setup for X-ray CT 
is shown in Fig. \ref{fig:mainfig} (a).
Here, we expose the object(s)\footnote{Object may refer to either a scene, a single object, or multiple objects scanned using X-ray CT.} to polychromatic 
X-ray radiation 
and images of the transmitted X-ray intensity 
are acquired by an energy-integrating detector 
at several view angles of the object as it is rotated.
The detector is a panel of sensor pixels 
that are arranged in the form of a two-dimensional matrix.
While our approach is broadly applicable to 
a wide range of X-ray imaging systems,
we focus our discussion specifically on cone-beam and 
parallel-beam X-ray system geometries. 
In cone-beam geometry, X-rays are modeled as
originating from an infinitesimally small X-ray point source.
Alternatively, X-rays are modeled as parallel rays
when propagating from the source to the detector
in parallel-beam geometry.
Cone-beam geometries are more challenging for reconstruction
since X-rays diverge in three dimensions.

The X-ray intensities measured at each detector pixel 
are a function of the total X-ray attenuation by the 
material along the X-ray path connecting the pixel 
and the source for that view angle.
4D reconstruction refers to estimating the 
spatial and temporal distribution of X-ray attenuation by the object. 
Since we image with polychromatic X-rays, 
we reconstruct energy averaged linear 
attenuation coefficients (LACs). 
The reconstructed LAC is a function of the
4D continuous coordinates $r=(t,z,y,x)$, 
where $t$ is the time, 
and $(z,y,x)$ are the 3D spatial coordinates.
Let $p_i$ denote the $i^{th}$ projection such 
that $0 \leq i < MN$, where $M$ is the total number 
of view angles and $N$ is the number of detector pixels.
Given a projection index $i$, we can compute its 
view index as $m=\lfloor\frac{i}{N}\rfloor$ 
and detector pixel index as $n=\left(i \mod\left(N\right)\right)$
such that $i=mN+n$.
The projection $p_i$ is proportional to the negative logarithm of the 
X-ray intensity measurement by a detector pixel.
If $\lambda_i$ and $\bar{\lambda}_i$ are the intensity measurements
with and without the object respectively,
then the $i^{th}$ projection is 
$p_i=-\log\left(\lambda_i/\bar{\lambda}_i\right)$.

For any projection $p_i$ shown in Fig. \ref{fig:mainfig} (b),
we estimate its value $\bar{p}_i$ as the weighted sum of 
the LAC, $\mu(r)$, at time $t_i$ and 
spatial coordinates contained within the orange-colored 
pyramidal volume that connects the X-ray source
to the detector pixel that measures $p_i$.
Here, $t_{i}$ is the measurement time instant for $p_i$.
We will denote the angle of rotation for the object
during acquisition of the projection $p_i$ as $\theta_i$.
Since the object is also assumed to lie inside a cylinder-shaped 
field of view (FOV), only the coordinates within this FOV
contribute to the estimate $\bar{p}_i$.
The FOV is a function of the center coordinate of
rotation and does not change with $\theta_i$.
Henceforth, we will refer to the intersection
of the pyramidal volume and the FOV as the 
region-of-influence for $p_i$.
Let $r_{i,j}=(t_{i},z_{i,j},y_{i,j},x_{i,j})$ 
be a coordinate sample such that $(z_{i,j},y_{i,j},x_{i,j})$ is 
contained in the region-of-influence for $p_i$.
We refer to the computation of $\bar{p}_i$ from 
the LAC values, $\mu(r_{i,j})$, at several coordinates
samples, $r_{i,j}\, \forall j$, as the forward model.
We use a ray-sampling method for generating the coordinate 
samples, $(z_{i,j},y_{i,j},x_{i,j})$, that is described 
in appendix \ref{sec:linproj}.

We account for rotation of the object during the CT scan
by rotation of the $x$ and $y$ axial coordinate system 
around the rotation axis center.
Rotation does not modify the $z$-axis coordinates 
since the rotation axis is parallel to the $z$-axis
and perpendicular to the $x-y$ axial plane.
If the object is rotated by $\theta_i$, 
it is equivalent to rotation of the source-detector pair 
by $(-\theta_i)$ around the same axis. 
Hence, we can use the approach in section \ref{sec:linproj}
to compute $\bar{p}_i$ after rotation of the source-detector pair by $(-\theta_i)$.
Alternatively, we can first generate coordinate samples using 
the method in appendix \ref{sec:linproj} 
while assuming $0^\degree$ of rotation.
Next, we compute $r_{i,j}\, \forall j$ 
by rotating the generated coordinates by $(-\theta_i)$ in the $x-y$ plane.
The time coordinate $t_i$ is solely dependent on the 
acquisition time of each projection image.
For the $m^{th}$ projection image, 
all pixels have the same acquisition time of $T_m$.
Hence, $t_i = T_m$ for all $(m-1)N \leq i < mN$.

To train the DINR network, 
we need a forward model that is a 
function of the outputs from the DINR network. 
The output of DINR, denoted by 
$\mathcal{M}(r_{i,j}; \gamma)$, 
is a measure of the LAC at 
the object coordinate $r_{i,j}$.
The neural network architecture for 
$\mathcal{M}(r_{i,j}; \gamma)$ is described 
in appendix \ref{sec:netarch}.
\textcolor{cgcol}{For $\mathcal{M}(r_{i,j}; \gamma)$, we chose
a fully connected neural network (FCNN)
with Fourier Feature Encoding from the paper \cite{Fourier_INR_NEURIPS2020}.}
Let $\Phi_i$ be the set of all indices $j$ 
for the samples of object coordinates $r_{i,j}$
within the region-of-influence for projection $p_i$.
The forward model as a function of 
$\mathcal{M}(r_{i,j}; \gamma)$ is,
\begin{equation}
    \label{eq:forwmodproj}
    \bar{p}_i = \frac{1}{\vert\Phi_i\vert}\sum_{j\in\Phi_i} 
w(r_{i,j}) \mathcal{M}\left(r_{i,j}; \gamma\right).
\end{equation}
Here, $w(r_{i,j})$ is a weight parameter 
that is a function of the spatial coordinates $(z_{i,j},y_{i,j},x_{i,j})$
and the imaging geometry. 
The formula for computation of $w(r_{i,j})$ 
is presented in appendix \ref{sec:linproj}. 
$w(r_{i,j})$ has two distinct purposes that is 
described by factorizing $w(r_{i,j})$ as,
\begin{equation}
\label{eq:weightfactors}
w(r_{i,j})=\mu_0\l(r_{i,j}).    
\end{equation}
The first term $\mu_0$ is used to scale the output 
$\mathcal{M}(r_{i,j}; \gamma)$ of DINR to have units of LAC.
The purpose of $\mu_0$ is also to improve the convergence
speed of the training optimization loop.
It is set to be approximately equal 
to the average value of the object's LAC.
The second term $l(r_{i,j})$ is another scaling factor 
for the output $\mathcal{M}(r_{i,j}; \gamma)$
that is equal to the
length of the ray containing $r_{i,j}$ 
inside the cylindrical FOV.
The expression $\frac{1}{\vert\Phi_i\vert}\sum_{j\in\Phi_i} w(r_{i,j}) \mathcal{M}(r_{i,j}; \gamma)$
is the unit-less linear projection of the LAC.  

We train a DINR to reconstruct the LAC as a 
function of the object coordinates by minimizing a loss function
that is a measure of the distance between the projection data
and its predictions using the forward model.
The network parameters, denoted by $\gamma$, are iteratively
optimized such that the projection estimate $\bar{p}_i$ is progressively
driven closer to the measured projection $p_i$ for every $i$.
We define a L2 loss function that is a 
measure of the squared distance between $p_i$
and its estimate $\bar{p}_i$ as,
\begin{equation}
    \label{eq:mainsqdist}
    d_i(\gamma) = \left(p_i-\bar{p}_i\right)^2 
    = \left(p_i-\frac{1}{\vert\Phi_i\vert}\sum_{j\in\Phi_i}  w(r_{i,j})\mathcal{M}(r_{i,j}; \gamma)\right)^2.
\end{equation}
Then, we train our DINR by estimating the $\gamma$ that
solves the following optimization problem,
\begin{equation}
    \label{eq:mainoptfunc}
    \hat{\gamma} = \arg\min_{\gamma} \frac{1}{MN}\sum_{i=1}^{MN} d_i(\gamma).
\end{equation}

Due to the computational and memory intensive nature of the 
minimization problem in equation \eqref{eq:mainoptfunc}, 
we solve it using distributed and stochastic optimization
over several compute processes.
The processes run on several GPUs distributed over
multiple compute nodes such that they collaboratively 
train the DINR to reconstruct the object.
Fig. \ref{fig:mainfig} (d) is a schematic of our approach to 
distributed optimization for learning of the 
network parameters $\gamma$.
In each process $k$, we randomly choose a subset $\Omega_k$
of projection indices $i$ from the set of $NM$ projection pixel indices.
As an example, the red-colored pixels in Fig. \ref{fig:mainfig} (c)
show the randomly chosen projection pixels whose indices
are contained in the set $\Omega_k$.
Then, we compute the local loss function 
$L\left(\Omega_k; \gamma\right)$ at each process $k$ 
such that it only sums the squared distances in equation \eqref{eq:mainsqdist}
over the small subset of indices $i\in\Omega_k$. 
The local loss function for each process is given by,
\begin{equation}
\label{eq:localoptfunc}
L\left(\Omega_k; \gamma\right)=
\frac{1}{\left\vert\Omega_k\right\vert}\sum_{i\in \Omega_k} d_i(\gamma),
\end{equation}
where $\vert\Omega_k\vert$ is the cardinality 
(number of elements) of the set $\Omega_k$.
Note that $\vert\Omega_k\vert$ is the batch size 
for process $k$ and is the same for all processes. 
Thus, if $K$ is the total number of processes, then 
the total batch size per iteration is,
\begin{equation}
\label{eq:totbatch}
\Omega_{*} = K\vert\Omega_k\vert.
\end{equation}
The batch size per GPU, $\vert\Omega_k\vert$, is adjusted
based on the amount of memory in each GPU.

We use back-propagation to compute the local gradient, 
$\nabla_{\gamma}L\left(\Omega_k; \gamma\right)$, 
of $L\left(\Omega_k; \gamma\right)$ with respect to $\gamma$ in each compute process $k$ (Fig. \ref{fig:mainfig} (d)).
This local gradient function
$\nabla_{\gamma}L\left(\Omega_k; \gamma\right)$
includes only the back-propagation of the loss terms 
$d_i(\gamma)\,\forall i\in\Omega_k$. 
Then, the gradients from all the processes are averaged and 
the averaged gradient is broadcast back to each individual process $k$.
This averaged gradient is denoted as 
$\nabla_{\gamma}L\left(\Omega_*; \gamma\right)$,
where the set $\Omega_*$ includes the projection indices 
from all the subsets $\Omega_k \forall k$. 
Finally, the optimizer updates the parameters $\gamma$ 
within each process.
We use the Adam optimizer \cite{adam_opt_2015} for estimation of $\gamma$.
Our approach to distributed optimization is highly scalable
where distinct processes are run simultaneously 
on several hundreds of distinct GPUs.
Each process maintains a local copy of the DINR network 
$\mathcal{M}\left(r; \gamma\right)$ 
and uses its own instance of the optimizer for updating $\gamma$.
Since the optimizer in each process uses the same averaged gradient, 
the parameter $\gamma$ after the update will also remain the same in all processes. 
One training epoch is the number of iterations
that is required to iterate over $MN$ number of projection pixels.
Thus, the number of iterations in each epoch 
is $\lceil\frac{MN}{\vert \Omega_{*}\vert}\rceil$
where $\Omega_{*}$ is the total batch size from equation \eqref{eq:totbatch}.
The parameters for training of 
$\mathcal{M}\left(r; \gamma\right)$
are described in appendix \ref{sec:trainpars}.
\textcolor{cgcol}{In appendix \ref{sec:trainpars}, we also
describe our approach to tuning the regularization parameters
that control the smoothness and continuity across space-time. 
We also present suitable default values that produce reasonable reconstructions for our tested datasets.
}

% \textcolor{cgcol}{Unlike explicit motion models \cite{Eyndhoven_2015_TIP, Boigne_2022_TCI, Nikitin_2019_TCI}
% or statistical prior models for space/time modeling \cite{mohan2015timbir, Kazantsev_2015_InvProblems} 
% during iterative reconstruction,
% INRs implicitly assign a low likelihood to spurious reconstruction artifacts and a higher likelihood to
% realistic image features.
% This property of neural networks is partly responsible for the recent tremendous boom
% in generative modeling of real world images and natural languages.
% Note that statistical prior models typically produce the highest likelihood
% for a zero valued reconstruction while also permitting a higher likelihood for noise 
% compared to the true object.
% While the analytical form of the motion models may avoid noise and artifacts 
% along the time axis, they do not explicitly mitigate them along the spatial dimensions.
% Since INRs are neural networks whose outputs are 
% continuous functions of the input coordinates,
% they tend to produce outputs that vary smoothly with the coordinates.
% Thus, INRs are inherently averse to reconstruction of noise
% and streak artifacts that typically plague 4DCT reconstructions.}

\section{Results}

We evaluate DINR on both experimental and simulated datasets. 
For the experimental data, we used a Zeiss Xradia 510 Versa 
(Carl Zeiss X-ray Microscopy, Inc., USA) X-ray imaging system to perform 
in situ 4DCT acquisitions of two samples under compression. 
A Deben CT5000 in-situ loadcell testing stage 
(Deben UK Ltd., United Kingdom) was used to 
compress the samples.
We performed in-situ 4DCT of crack propagation
in a SiC cylindrical sample and feature deformation
in a polydimethylsiloxane ``log pile'' sample \cite{osti_1871381}. 
\textcolor{cgcol}{We made these real experimental dynamic 4DCT datasets publicly available \cite{DebenDataset}}. 
For ablation studies and 
a more rigorous quantitative evaluation, 
we used LLNL D4DCT dataset \cite{MPMDataset}, a simulated dataset 
for reconstruction of object deformation under mechanical loads over time. 
For simulation of CT projections, we used Livermore Tomography Tools (LTT) \cite{LTT} 
and Livermore AI Projector (LEAP) \cite{LEAPCT}. 
We provide image quality metrics 
for quantitative evaluation of the reconstruction as well as
comparison with SOTA methods. 

\subsection{Experimental Data Evaluation}

\begin{table*}
    \centering
    \begin{tabular}{|c|c|c|}
    \hline
          & Log-pile sample & SiC sample \\
    \hline
        Detector pixel size & 69.16 $\mu m$ &  69.16 $\mu m$ \\
    \hline
        Object voxel size (High resolution DINR) & 39.52 $\mu m$ & 13.83 $\mu m$ \\
    \hline 
        Source to object distance (SOD) & 80 $mm$ & 65 $mm$ \\ 
    \hline
        Source to detector distance (SDD) & 140 $mm$ & 325 $mm$ \\
    \hline 
        Number of view angles & 722 & 200 \\
    \hline
      Total rotation angle over all views & 722$^{\circ}$ & 397.8$^{\circ}$ \\ 
    \hline
        Shape of projections & 722 $\times$ 400 $\times$ 1024 & 200 $\times$ 328 $\times$ 1024 \\
    \hline 
        Shape of volume during inference (High Resolution DINR) & 722 $\times$ 400 $\times$ 1024 $\times$ 1024 & 200 $\times$ 328 $\times$ 1024 $\times$ 1024\\
    \hline 
        Comparison figures & Fig. \ref{fig:deben202303} & Fig. \ref{fig:deben202310}\\
    \hline
    \end{tabular}
    \vspace{0.02in}
    \caption{\textcolor{cgcol}{Geometric and reconstruction parameters for the CT experimental datasets of the log-pile and SiC samples.}}
    \label{tab:exp_params}
\end{table*}

\begin{figure*}[!htb]
\begin{center}
\begin{tabular}{cccc}
\includegraphics[height=1.2in]{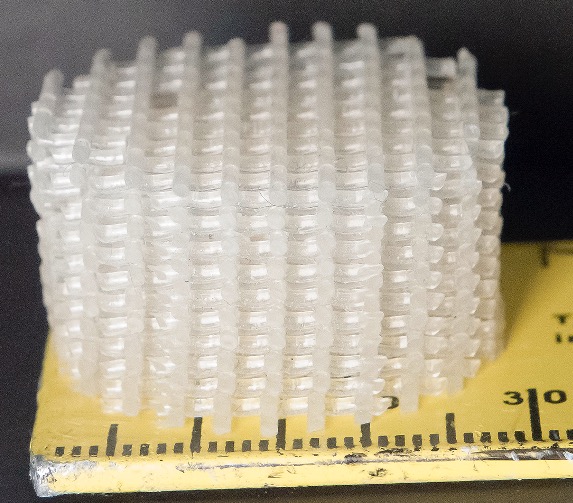} &
\includegraphics[height=1in]{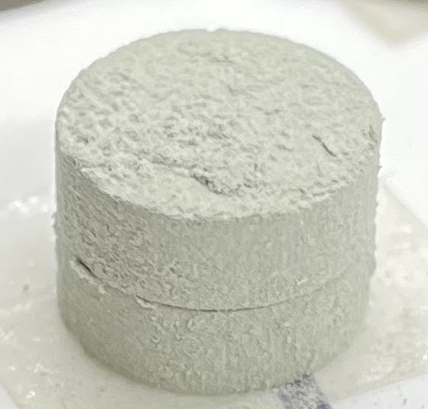} &
\includegraphics[height=1in]{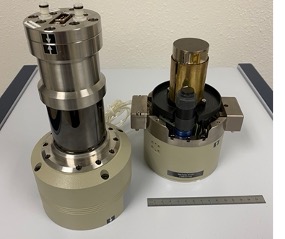} &
\includegraphics[height=1in]{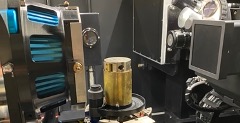} \\
(a) Log-pile & (b) SiC & (c) Deben stage & (d) Zeiss Xradia 510 Versa \vspace{0.05in}\\
\end{tabular}
\includegraphics[width=6in]{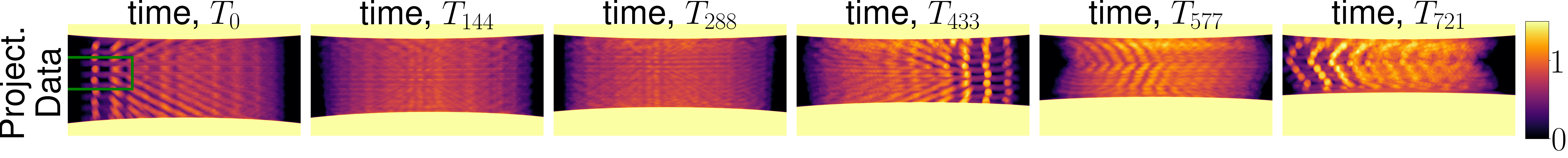}\\
\includegraphics[width=6in]{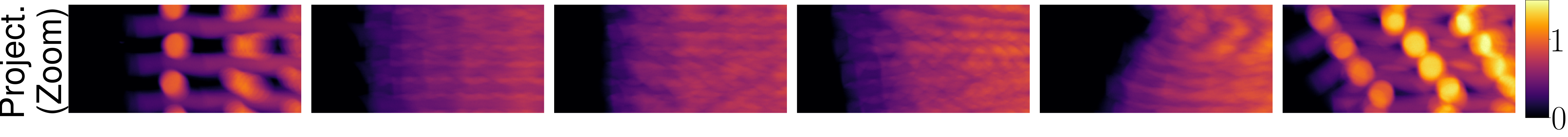}\\
\vspace{-0.1in}
(e) X-ray projection images of the log pile sample.\vspace{0.05in}\\
\includegraphics[width=6in]{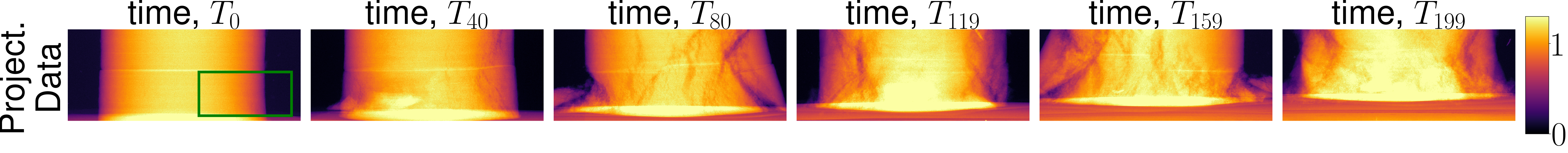}\\
\includegraphics[width=6in]{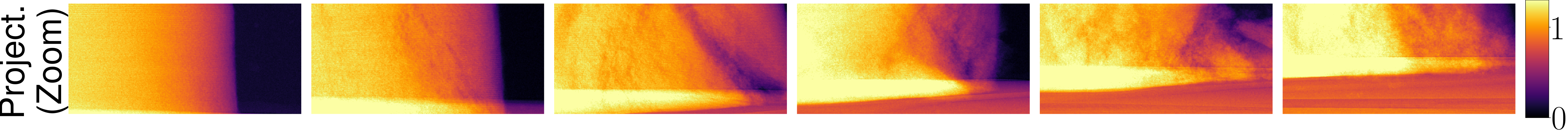}\\
\vspace{-0.1in}
(f) X-ray projection images of the SiC sample.\\
\end{center}
\caption{\label{fig:debendata}
4DCT of samples under compression in a Deben stage
that is mounted in a cone-beam X-ray CT system.
(a) and (b) show the log-pile and SiC samples, respectively, 
used for the 4DCT scans. 
(c) shows the Deben stage used for in-situ compression
of the samples.
(d) shows the Zeiss Xradia 510 Versa cone-beam X-ray 
imaging system used for 4DCT acquisitions.
(e) and (f) show the X-ray projection images 
of (a) and (b) respectively at different view angles. 
$T_m$ in the column labels of (e, f)
indicates the time of the projection images 
at the $m^{th}$ view.
In (e, f), the $2^{nd}$ row shows magnified views 
of the first-row images in the region denoted 
by the rectangular box at time $T_0$.
}
\end{figure*}

%The stage is contained within a cylindrical carbon fiber envelope, which must be considered when performing flat field correction as the envelope contributes to X-ray attenuation.
A polydimethylsiloxane ``log-pile'' sample additively manufactured 
using direct ink writing (DIW) is shown in  Fig. \ref{fig:debendata} (a).
The log pile sample is comprised of several layers of strands;
between adjacent layers, the strands are rotated by $90^{\circ}$.
The distance between adjacent layers is approximately $0.750$ mm.
%Separating the siloxane from the steel anvils of the Deben stage 
%are disks of soft plastic that are approximately $1$ mm in thickness.
The Silicon Carbide (SiC) sample shown in Fig. \ref{fig:debendata} (b) 
consists of two vertically stacked Silicon Carbide (SiC)
cylinders (or ``pucks'') made by binder jetting - 
another additive manufacturing technique. 
Each puck was approximately $8$ mm in diameter 
and approximately $2.8$ mm in height.
%Separating the pucks from the steel anvils on either side 
%of the Deben stage are $2$-mm thick plastic disks. 

\newcommand{\debmarhsp}{\hspace{-0.15in}}
\makeatletter
\define@key{Gin}{debmarfig}[true]{%
    \edef\@tempa{{Gin}{width=0.7in}}%
    \expandafter\setkeys\@tempa
}
\makeatother
\makeatletter
\define@key{Gin}{debmariso}[true]{%
    % https://tex.stackexchange.com/questions/57418/crop-an-inserted-image
    % Answer: [trim={left bottom right top},clip]
    %\edef\@tempa{{Gin}{trim={150px 350px 150px 350px},clip,width=0.8in,height=0.7in}}%
    \edef\@tempa{{Gin}{width=0.8in,height=0.8in}}%
    \expandafter\setkeys\@tempa
}
\makeatother

\begin{figure*}[!htb]
\begin{center}
\hspace{-0.1in}
\includegraphics[width=6in]{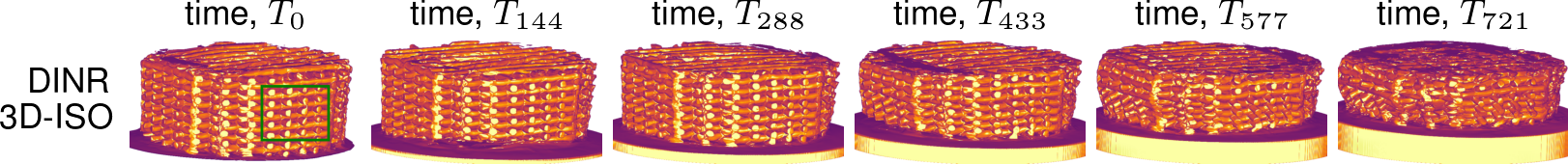} 
\vspace{0.05in}\\
\hspace{-0.1in}
\includegraphics[width=6in]{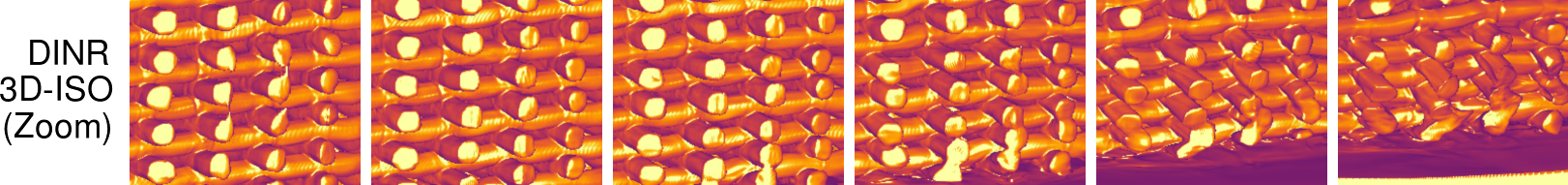} \vspace{0.05in}\\
\hspace{-0.05in}
\includegraphics[width=5.9in]{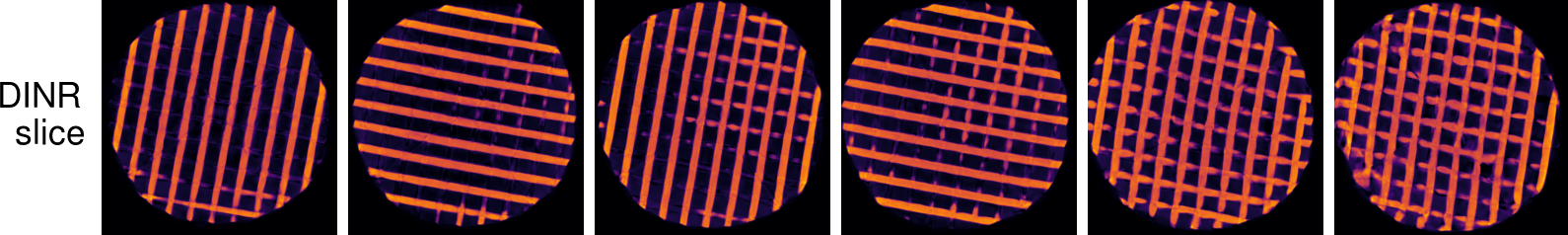} \vspace{-0.05in}\\
(a) Sequence of time frames reconstructed using DINR.\vspace{0.05in}
\includegraphics[width=6.5in]{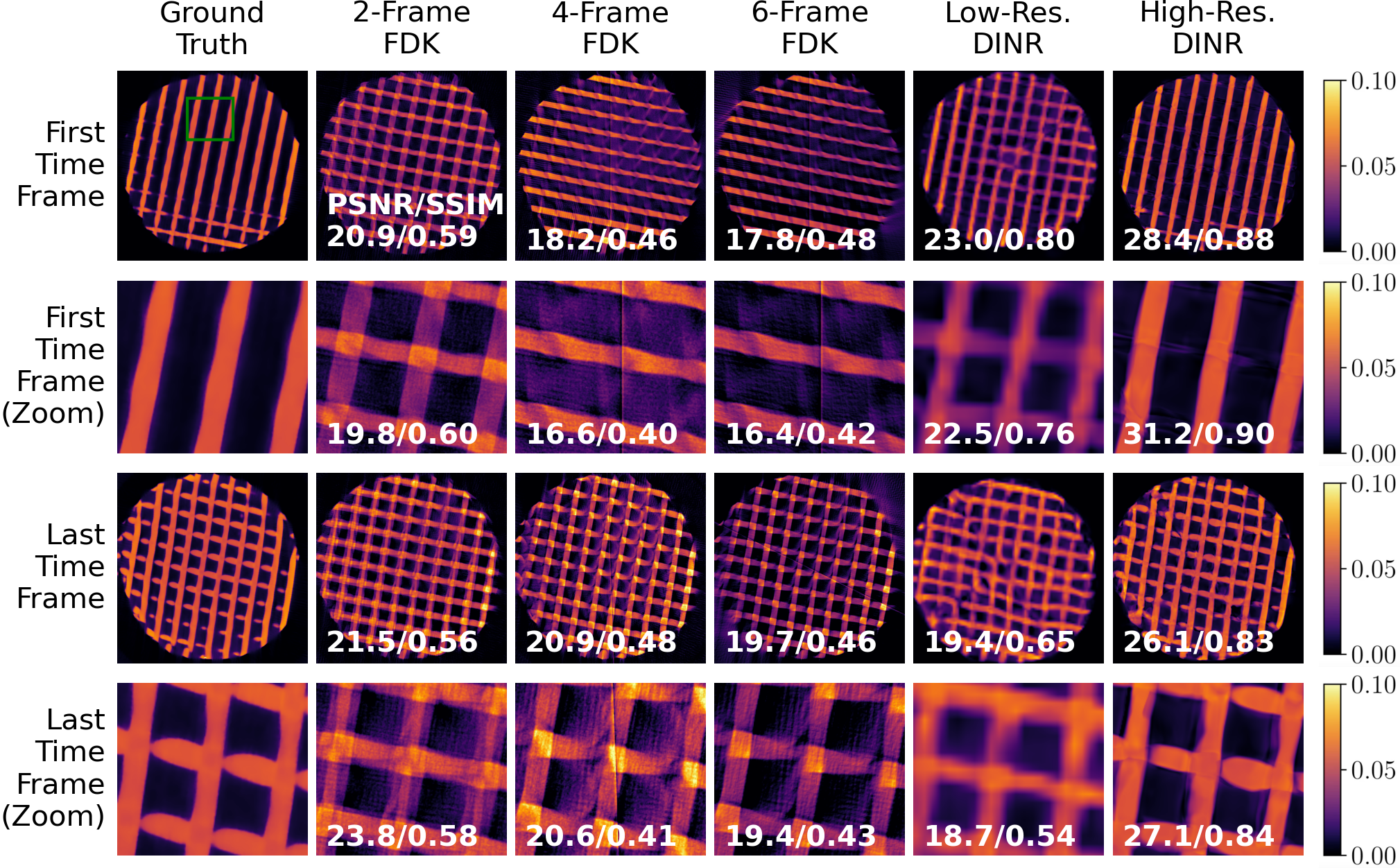} \\
(b) Comparison between 
FDK (conventional 4DCT) and DINR (ours). \\
\end{center}
\caption{\label{fig:deben202303}
4DCT reconstruction of the log-pile sample
(Fig. \ref{fig:debendata} (a)).
(a) shows the 3D ISO surface of the 4D reconstruction
and the cross-section images of the LAC 
using the high-resolution DINR at various times. 
(b) is a reconstruction comparison of the 
cross-section images of the LAC between the 
conventional 4D FDK and our DINR approach.
The PSNR/SSIM values are embedded 
in the images of (b).
The high-resolution DINR images are the best
visual match for the ground-truth while also
producing the highest PSNR and SSIM.
\textcolor{cgcol}{Fig. \ref{fig:log-pile-fdk-rwls-dinr}
in the supplementary document  
demonstrates the advantage of
DINR compared to the Regularized
Weighted Least Squares (RWLS) algorithm \cite{LEAPCT}
with total variation regularization in 3D.}
}
\end{figure*}

%\makeatother
%\makeatletter
%\define@key{Gin}{deboctiso}[true]{%
%    % https://tex.stackexchange.com/questions/57418/crop-an-inserted-image
%    % Answer: [trim={left bottom right top},clip]
%    \edef\@tempa{{Gin}{trim={200px 200px 200px 200px},clip,width=0.8in,height=0.7in}}%
%    \expandafter\setkeys\@tempa
%}
%\makeatother
\begin{figure*}[!thb]
\begin{center}
\includegraphics[width=6in]{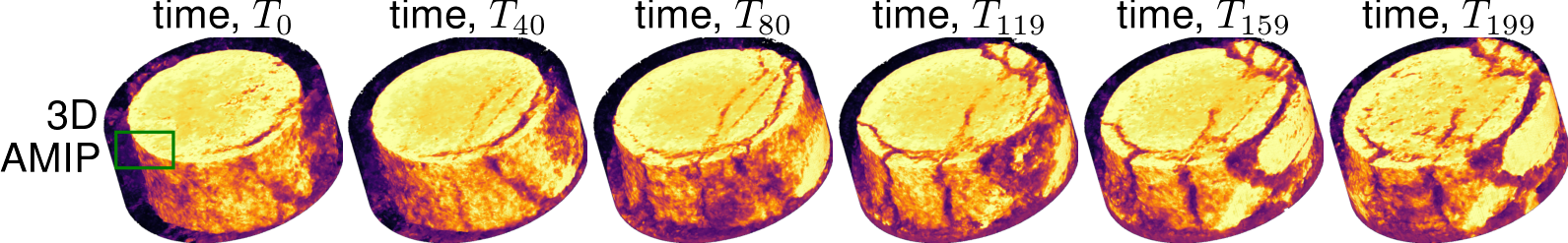}\vspace{0.05in}\\
\includegraphics[width=6in]{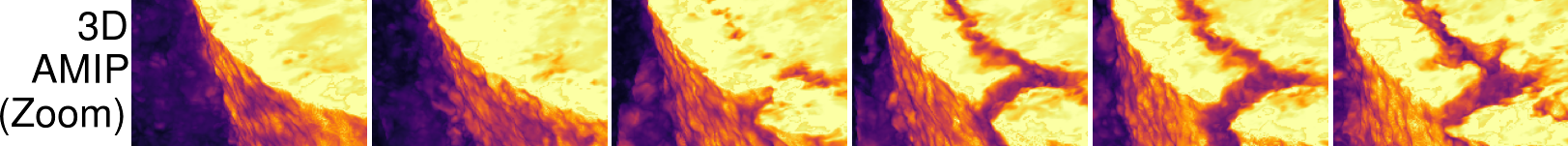}\\
(a) 3D renderings of the high-resolution DINR reconstruction. \\
\vspace{0.05in}
\hspace{-0.2in}
\includegraphics[width=6.5in]{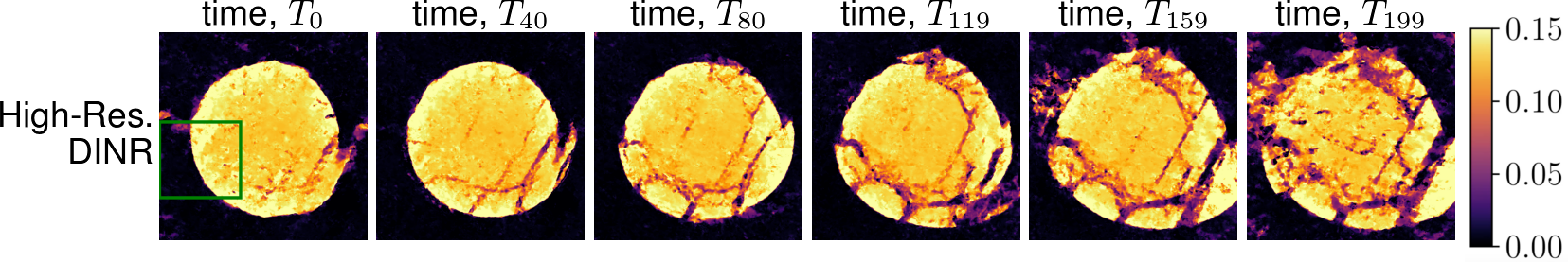}\vspace{0.05in}\\
\vspace{-0.05in}
\hspace{-0.2in}
\includegraphics[width=6.5in]{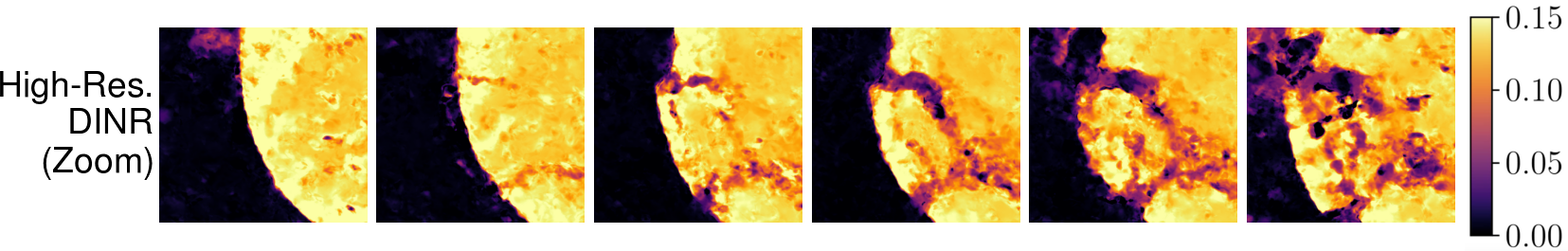}\vspace{0.05in}\\
\vspace{-0.07in}
\hspace{-0.15in}
\includegraphics[width=6.5in]{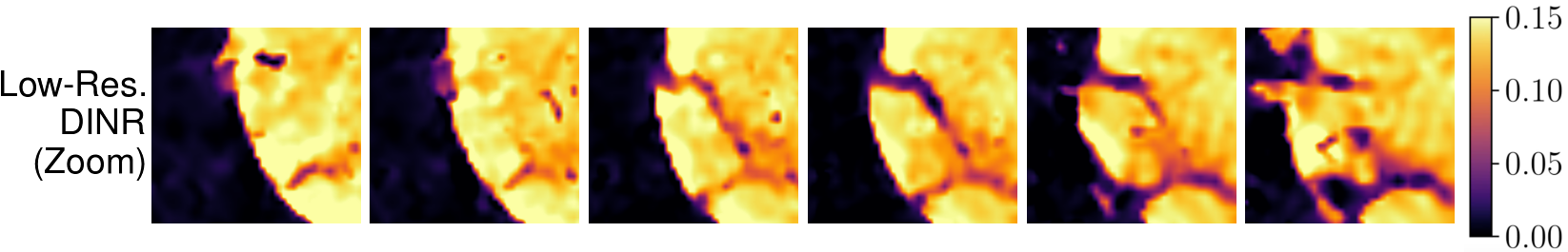}\\
\vspace{-0.1in}
(b) 4DCT reconstructions of the LAC using DINR.
\vspace{0.05in} \\
\includegraphics[width=6.5in]{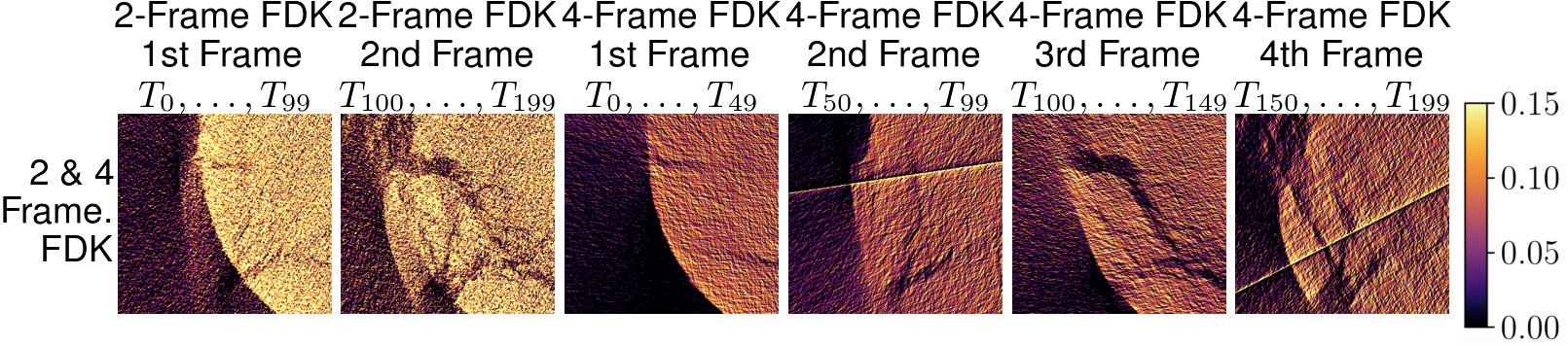}\\
\vspace{-0.1in}
(c) Conventional 2-frame and 4-frame FDK 4DCT reconstruction.
\\
\end{center}
\caption{\label{fig:deben202310}
4DCT reconstruction of the SiC sample (Fig. \ref{fig:debendata} (b)).
(a) shows the 3D AMIP volumes of the high resolution 
4D DINR reconstruction with clearly resolved 
propagation of cracks over time. 
(b) shows the high and low resolution DINR reconstruction
of a cross-axial slice at different times.
The low resolution DINR is unable to clearly resolve the cracks.
The high resolution DINR produces the best reconstruction that
clearly resolves the cracks in all cases.
The first two images and the last four images in
(c) show the 2-frame FDK and the 4-frame FDK 
reconstructions respectively.
The 2-frame FDK suffers from motion blur while 
the 4-frame FDK has substantial limited angle artifacts.
}
\end{figure*}

\begin{figure}[hbtp]
    \includegraphics[width=0.8\linewidth]{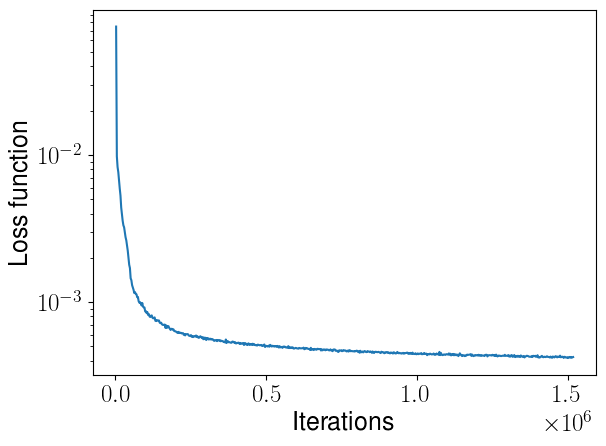}
\caption{\textcolor{cgcol}{Plot of the total loss function, $L\left(\Omega_*, \gamma\right)$, vs. iterations for the log-pile experimental data reconstruction using 
High-Res. DINR. Each iteration is approximately
$0.214$ seconds when training on $128$ Nvidia V100 GPUs. We observe that the convergence of the loss function
is highly stable. The time for one 3D inference, i.e., the 
time taken for a single 3D volume reconstruction 
(Table \ref{tab:exp_params}), is approximately $38$ minutes using $48$ GPUs.}}
\label{fig:deb202303_loss}
\end{figure}

\begin{figure*}[hbtp]
\centering

\subfloat[Reconstruction of first (top row) and last (bottom row) time frames for S08\_005\label{fig:mpm1}]{
       \includegraphics[width=0.9\linewidth]{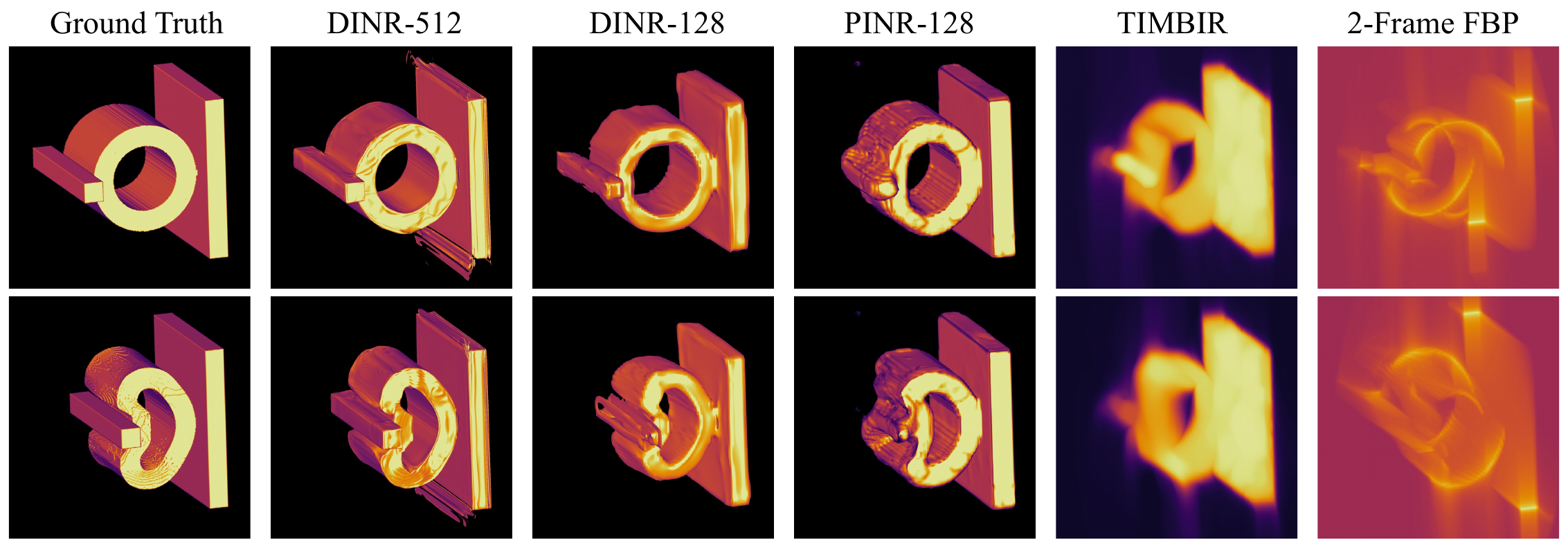}}

\subfloat[Reconstruction of first (top row) and last (bottom row) time frames for S04\_018\label{fig:mpm2}]{
       \includegraphics[width=0.9\linewidth]{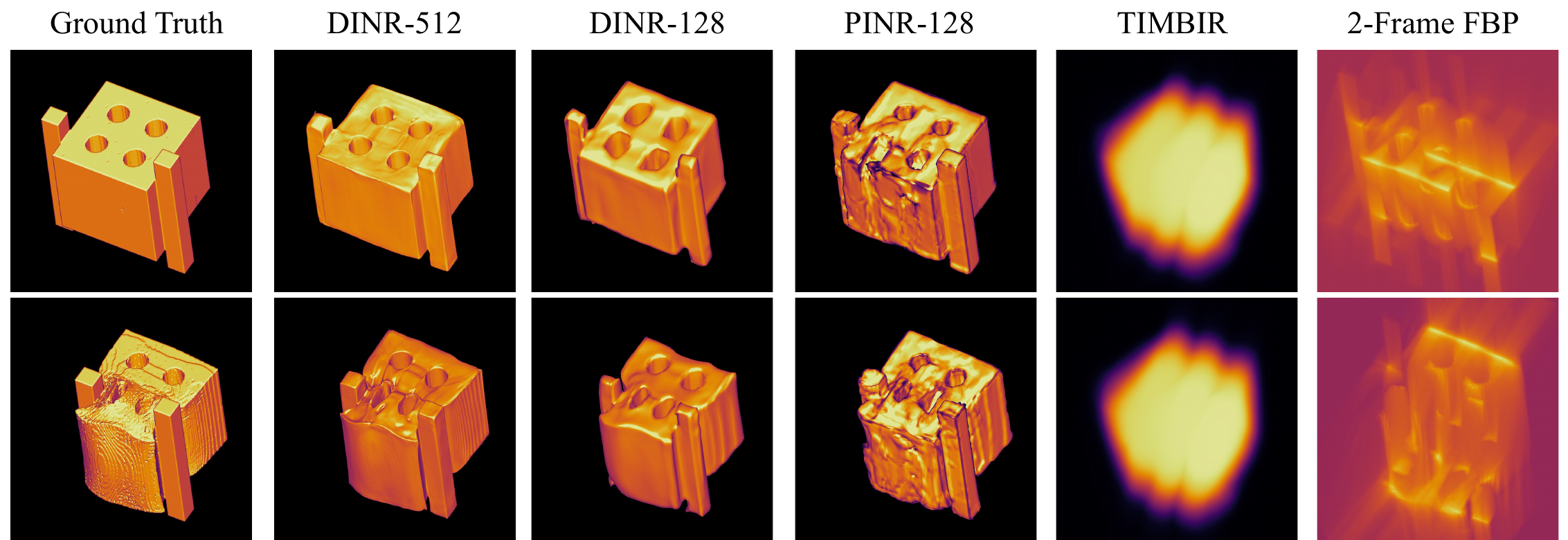}}

\caption{Qualitative (visual) comparison of 4DCT reconstructions of our simulated MPM datasets between DINR and existing SOTA methods. (a) and (b) show the 4D reconstructions 
for the S08\_005 and S04\_018 MPM datasets. 
%The $1^{st}$ row of (a, b) show the first time frame.
%The $2^{nd}$ row of (a, b) show the last time frame. 
We used isosurface 3D visualization for DINR and PINR.  
For TIMBIR and 2-Frame FBP,  we instead used MIP 
volume rendering due to unintelligible isosurfaces.}
\label{fig:mpm_qual}
\end{figure*}

\begin{figure*}[hbtp]
\centering

\begin{tabular}[b]{c}
    \includegraphics[width=.5\linewidth]{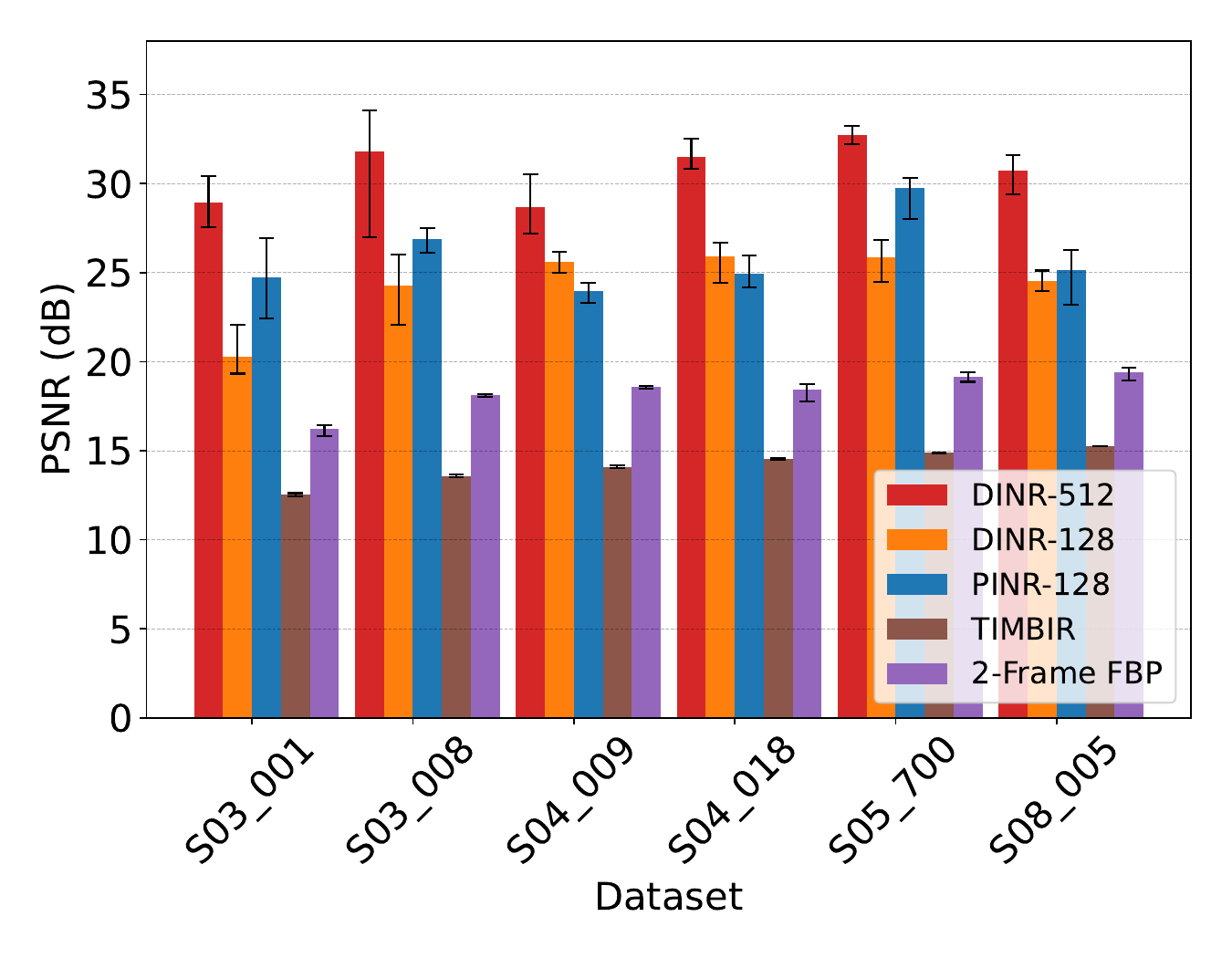} \\
    (a) Average PSNRs
    \label{fig:mpm_psnr}
\end{tabular} \qquad
\begin{tabular}[b]{c}
    \includegraphics[width=.4\linewidth]{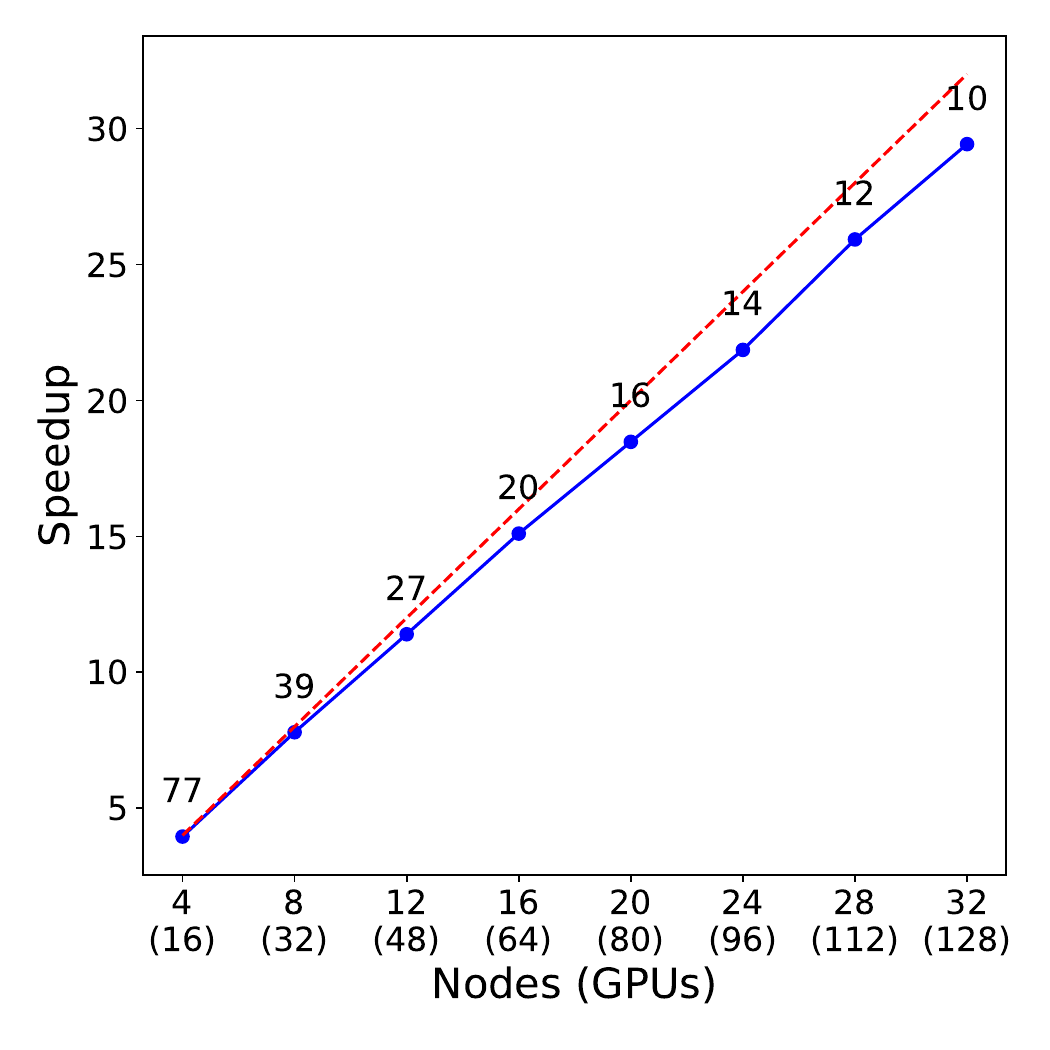} \\
    \textcolor{cgcol}{(b) Strong Scaling}
    \label{fig:mpm_scaling}
\end{tabular}
  
%\subfloat[Average PSNRs\label{fig:mpm_psnr}]{
%       \includegraphics[width=0.4\linewidth]%{figs/figure_MPM_512_128_psnr.pdf}}
%}
%\subfloat[Strong Scaling\label{fig:mpm_scaling}]{
%       \includegraphics[width=0.4\linewidth]{figs/figure_MPM_strong_scaling2.pdf}}

\caption{Quantitative comparison of 4DCT reconstructions of our simulated MPM datasets between DINR and existing SOTA methods. 
(a) shows the average PSNRs between the reconstructions and ground-truth. The error bars in (a) indicate the minimum and maximum PSNRs. \textcolor{cgcol}{DINR-512 has the highest PSNRs. While the PSNR comparison between PINR-128 and DINR-128 is inconclusive, the reconstructed images in Fig. \ref{fig:mpm_qual} clearly demonstrate the superior reconstruction of DINR-128 compared to PINR-128.} (b) \textcolor{cgcol}{illustrates the strong scaling of DINR-512 in relation to the number of compute nodes and GPUs. Each point is labeled (above the blue curve) with its corresponding training time in minutes.} }
\label{fig:mpm_quant}

\end{figure*}

Each sample is individually mounted in the Deben testing stage shown in   Fig. \ref{fig:debendata} (c).
We used plastic disk spacers to separate the samples
from the steel anvils of the Deben stage.
The Deben stage and the loaded sample are then 
mounted in the X-ray imaging system, the interior of which is 
shown in Fig. \ref{fig:debendata} (d).
During acquisition, 
each sample was compressed at a speed of $0.030$ mm/min; 
the lower anvil of the testing stage applies compression 
by moving upward, while the upper anvil remains stationary.
\textcolor{cgcol}{The X-ray CT scans need calibration to remove the impact
of the attenuation caused by the Deben stage's envelope
that surrounds the object. 
We described this calibration procedure in the 
supplementary document.}
The exposure time for each X-ray projection image was $10$ seconds.
Thus, the acquisition time for the $m^{th}$
projection in units of seconds is,
\begin{equation}
\label{eq:projtime}
T_m \approx 10m.    
\end{equation}
The dynamic nature of the scene means that each projection image
captures a different sample morphology when 
compared to the previous projection image. 
Since the sample is also continuously rotating,
each projection image captures a different view angle
of the sample.

In Fig. \ref{fig:debendata} (e), 
we show projection images at different view angles
of the log-pile sample.
The second row shows magnified views of a rectangular 
region of the first-row images; 
the location of the region is shown 
in the first-row image corresponding to time $T_0$. 
The times and the corresponding indices for the projection images of 
Fig. \ref{fig:debendata} (e) are indicated by $T_m$
in the column labels.
We acquired $722$ X-ray projection images 
over two full rotations of $361^{\circ}$
and show the projection images at the time/view indices 
of $m = 0, 144, 288, 433, 577$ and $721$.
We observe buckling of the log-pile sample over time from different views
as the anvil of the Deben stage progressively 
compresses the sample.
%During each acquisition of $361$ projections 
%spanning $360$ degrees of rotation, 
%the sample was compressed by approximately $2$ mm. 
The sample was compressed by approximately $4$ mm 
over the duration of the experiment.
\textcolor{cgcol}{The anvil compresses the log-pile sample by $0.13$ pixels 
in the duration of one projection image acquisition.}

X-ray projection images of the SiC sample at different view angles
are shown in Fig. \ref{fig:debendata} (f); 
the second row shows magnified views of  the first-row images.
We acquired $200$ projection images over a total angular 
rotation of approximately $397.8$ degrees.
The total compression of the sample was approximately $1$ mm
over the duration of the $200$ views.
Fig. \ref{fig:debendata} (f) shows the projection images
at the time/view indices of $m = 0, 40, 80, 119, 159,$ and $199$.
We observe progressive crack propagation over time 
near the edges of the SiC sample in the projection images.
\textcolor{cgcol}{The anvil compresses the SiC sample by $0.36$ pixels 
in the duration of one projection image acquisition.
Please note that the motion or deformation in the samples can be substantially faster than the speed of the anvil. }

The conventional approach to 4DCT is to split the 
projection images into groups such that
each group of projections is reconstructed into one volumetric time frame.
The projections in each group are from consecutive 
views or time indices without any overlap of indices between groups.
This approach to 4DCT relies on an assumption
of static object for the time duration of the projections within each group.
However, since each group may span tens to hundreds of projection images,
we are only able to reconstruct samples with very slow dynamics
using this approach.
Feldkamp-Davis-Kress (FDK) \cite{Feldkamp84_FDKrecon}
is an analytical cone-beam reconstruction algorithm that is
used for experimental data comparisons.
Henceforth, we use the notation $K-$Frame FDK to refer to 
$K$ number of FDK reconstructed time frames 
(each time frame is a volumetric reconstruction) 
from $K$ groups of projections respectively.

If the static assumption for the object 
is violated within any group of projections, 
we observe substantial motion artifacts and blur
in the reconstructed volumes.  
If the number of views in each group is reduced
to mitigate motion blur,
we observe substantial limited angle artifacts (streaks)
when the angular span of rotation reduces to less than $180^{\circ}$
for parallel-beam geometry.
For cone-beam, this limit is a variable that depends on the cone-angle,
but is generally a higher number that varies 
between $180^{\circ}$ and $360^{\circ}$.
The time-interlaced model-based iterative
reconstruction (TIMBIR) algorithm \cite{mohan2015timbir} 
also suffers from the same limitation while also 
currently only supporting the parallel-beam CT geometry.
In contrast, DINR assumes a dynamic scene that allows temporal 
changes of the object between consecutive projection images.
This property of DINR enables the reconstruction 
of rapidly changing scenes.

In Fig. \ref{fig:deben202303}, we demonstrate that our high-resolution
DINR reconstruction is able to most accurately 
resolve the strands of the log-pile sample. 
Here, high-resolution refers to reconstruction of volumes
using projection images 
at a pixel size of $69.16\,\mu m$ 
without down-sampling. 
The high-resolution DINR reconstruction at various time indices
is shown in Fig. \ref{fig:deben202303} (a).
We show reconstructions at the same time indices as the
projection images in Fig. \ref{fig:debendata} (e).
The images in the various columns are at time
indices of $m = 0, 144, 288, 433, 577$ and $721$.
The $1^{st}$ row of Fig. \ref{fig:deben202303} (a) 
shows the 3D renderings of the isosurfaces 
(labeled as ISO in Fig. \ref{fig:deben202303} (a))
computed from the reconstructed volumes.
We used Napari \cite{napari_ahlers_2023} for generating isosurfaces.
Each ISO surface was produced from a volumetric reconstruction
comprising of $400 \times 1024 \times 1024$ voxels.
The $2^{nd}$ row of Fig. \ref{fig:deben202303} (a) 
shows a magnified view of the rectangular region of 
interest in the isosurface renderings along the $1^{st}$ row. 
The $3^{rd}$ row of Fig. \ref{fig:deben202303} (a) 
shows cross-axial slices of the reconstructed volumetric time frames.
These cross-axial images are in units of the 
linear attenuation coefficient (LAC)
given by the DINR output $\mathcal{M}(r_{i,j}; \gamma)$ multiplied by $\mu_0$
(from equation \eqref{eq:weightfactors}).
The DINR outputs are obtained by neural network inference
after training.

For comparison, 
we acquired static CT scans of the log-pile sample
prior to and after compression.
FDK reconstructions of the sample from 
these static CT scans are treated as ground truth
and shown in the $1^{st}$ column of Fig. \ref{fig:deben202303} (b).
Using these ground truth images, 
we compute the peak signal to noise ratio (PSNR) 
and structural similarity index (SSIM) metrics 
for assessing the performance of the 
various 4DCT reconstruction approaches. 
The PSNR (units of dB) and SSIM metrics are computed 
after min/max normalization
of the reconstructed images using the minimum and maximum values of the 
ground-truth images.

The conventional 2-Frame FDK in the $2^{nd}$ 
column of Fig. \ref{fig:deben202303} (b)
uses FDK reconstruction of the first $361$ projection images 
as the first frame and that of the last $361$ projection images
as the last frame respectively.
Similarly, the 4-Frame FDK in the $3^{rd}$ column shows FDK reconstructions 
of the first and the last $180$ number of projection images
as the first and last frames respectively.
The $2$-, $4$-, and $6$-Frame FDKs lead to very low fidelity
reconstructions with misplaced positioning of image features.
The dynamics or motion in the sample during the time span of each
frame for $2$-Frame FDK causes significant motion blur.
With $4$- and $6$-Frame FDK, in addition to blur, we observe 
limited angle artifacts since the angular rotation 
is approximately $180^{\circ}$ and $120^{\circ}$ respectively.

We evaluate DINR for 4D reconstruction of the log-pile sample
under compression for two cases of low and high resolution 4DCT.
The low-resolution DINR reconstructions 
in the $5^{th}$ column of Fig. \ref{fig:deben202303} (b) 
are from projection images downsampled by a factor of $8$.
Alternatively, the high-resolution DINR reconstructions
in the last column of Fig. \ref{fig:deben202303} (b) 
are from projection images without any down-sampling.
The advantage of the low-resolution DINR is the significantly lower
memory and compute costs due to the substantially smaller data size. 
However, this approach leads to severely blurred image features 
due to a $8\times$ increase in pixel size.
Prior neural representation approaches, e.g., \cite{Reed2021} 
require substantial down-sampling of projection data
to meet computational and memory requirements. 
Unfortunately, such down-sampling may not be suitable for cases
where high-resolution is a necessity.
Among all approaches, the high-resolution DINR provides
the best reconstruction fidelity 
and is the only method to accurately reconstruct
the strands of the log-pile sample.
The highest PSNR and SSIM is achieved by the high-resolution DINR
along each row of Fig. \ref{fig:deben202303} (b).
The pixel sizes of the projection images for the low
and high-resolution cases were $553.2\,\mu m$ and $69.16\,\mu m$, respectively.
\textcolor{cgcol}{We include a convergence plot for the loss function
as a function of the iterations in Fig. \ref{fig:deb202303_loss}. 
We also provide run times for each training iteration
and inference in the caption. }

We demonstrate the effective reconstruction of crack propagation 
inside a SiC sample using DINR in Fig. \ref{fig:deben202310}.
Fig. \ref{fig:deben202310} (a) show 
3D renderings using attenuated maximum intensity projection (AMIP)
of the volumetric reconstructions at different time steps.
We used Napari \cite{napari_ahlers_2023} to generate the AMIP 3D renderings.
We show time frames from the 4DCT reconstruction at 
projection indices of $m = 0, 40, 80, 119, 159, $ and $199$. 
The times for these reconstructions directly correspond
to the morphology of the sample during the acquisition times 
of the projection images shown in Fig. \ref{fig:debendata} (f).

We show the high and low resolution
DINR reconstructions of a cross-axial slice 
for the SiC sample in Fig. \ref{fig:deben202310} (b).
These reconstructions are in units of LAC
similar to Fig. \ref{fig:deben202303} (b). 
The low resolution DINR images in the $3^{rd}$ row of  
Fig. \ref{fig:deben202310} (b) are reconstructed from 
projection images that are down-sampled by a 
factor of $8$ along each image dimension.
Due to the lower resolution, the low resolution DINR is unable
to clearly resolve the cracks, especially at later time steps.
The high-resolution DINR reconstructions in the $1^{st}$ and $2^{nd}$
rows of Fig. \ref{fig:deben202310} (b) produce substantially
improved images that clearly resolve 
the cracks at all times.
In Fig. \ref{fig:deben202310} (c), we show the conventional 
$2$-Frame FDK and $4$-Frame FDK reconstruction 
of a cross-axial slice.
The $2$-Frame FDK shows reconstructed time frames from $2$ projection groups
each consisting of $100$ consecutive projection images.
The $4$-Frame FDK shows reconstructed time frames from $4$ projection groups,
each consisting of $50$ consecutive projection images.
None of the 4DCT reconstructions using FDK can clearly
resolve crack propagation due to insufficient temporal resolution.
$2$-Frame FDK suffers from motion blur due to substantial crack 
propagation over $100$ views.
$4$-Frame FDK produces limited angle artifacts
due to a rotation of approximately $99^{\circ}$
for the projections in each group.

\subsection{Simulated Data Evaluation}

We compare the performance of DINR to other SOTA 4DCT approaches 
on the reconstruction of simulated parallel-beam acquisitions 
of the LLNL D4DCT dataset ~\cite{MPMDataset}. 
This dataset was generated using the material point method (MPM)~\cite{MPM_CPDI2} to precisely simulate the deformation 
of a 6061 aluminum alloy object. 
Data was simulated for a range of object shapes and compression scenarios. 
For each dataset, the simulated volumes of object deformation over time 
are voxelated into $360$ time frames. 
These MPM datasets span a range of volume resolutions 
and each dataset includes a subset of the total time frames. 
Among $157$ datasets of varying object shapes and compression scenarios
in LLNL D4DCT Datasets~\cite{MPMDataset}, 
we selected $6$ for our comparisons: 
S03\_001, S03\_008, S04\_009, S04\_018, S05\_700, and S08\_005. 

We evaluate the performance of DINR and other SOTA 4DCT reconstruction 
approaches on two volumetric resolutions with pixel sizes 
of $2\,mm$ (low resolution) and $0.05\,mm$ 
(high resolution) in Fig. \ref{fig:mpm_qual}. 
For the lower resolution dataset ($2$mm pixel size), 
we simulated $91$ volumetric time frames comprising $128^3$ voxels.
For the higher resolution dataset ($0.05$mm pixel size), 
we simulated $181$ volumetric time frames comprising $512^3$ voxels. 
\textcolor{cgcol}{While the amount of motion and deformation varies depending on the data and the region within each data sample, the speed of movement in the lower resolution dataset can be up to $0.3$ pixels between consecutive projection views.}
Using the Livermore Tomography Tools (LTT) [30] software library, 
we simulated CT projection images from the time frames
such that the $m^{th}$ image was simulated from the $m^{th}$ time frame.
This ensures each projection image is generated from a 
unique volumetric time frame to simulate 4DCT of a 
continuously deforming object. 
The rotation angular range for both the low and high resolution
scenarios was $180^{\circ}$. 
The projection geometry is parallel-beam for CT data simulation. 
We also simulate $0.1\%$ Poisson noise in the X-ray transmission space,
i.e., the negative exponential of projections.
The reader is advised to consult the 
dataset website \cite{MPMDataset} for more details. 

We compare DINR against two SOTA 4DCT reconstruction methods:
TIMBIR~\cite{mohan2015timbir} and Parametric INR (PINR)~\cite{Reed2021}. 
We also performed limited-angle CT ($90^{\circ}$) reconstruction 
using 2-Frame filtered back projection (FBP). 
Figure~\ref{fig:mpm_qual} (a) and (b) shows 3D visualizations of 
the 4D reconstruction for S08\_005 and S04\_018 
using DINR, PINR, TIMBIR, and 2-Frame FBP. 
We used isosurface 3D visualization for DINR and PINR.
The presence of severe artifacts in the TIMBIR and FBP 
reconstructions renders isosurfaces unintelligible.
Hence, we visualized TIMBIR- and FBP-reconstructed 
volumes using maximum intensity projection (MIP)-based volume rendering.
DINR-128 and PINR-128 were reconstructed from the lower resolution 
projection data.
DINR-512 was reconstructed from the higher resolution
projection data.
Compared to the ground-truth, 
DINR-512 produces the best visual quality of reconstruction
for both S08\_005 and S04\_018.
Reconstructions using TIMBIR are almost unrecognizable
since the algorithm is optimized for sparse views 
with interleaved angular sampling, while this dataset
constitutes the more challenging limited-angle inverse problem. 
For TIMBIR, we carefully tuned the regularization parameters 
to obtain the best visual reconstruction quality. 
For more reconstructions of other datasets, 
please refer to the images in the supplementary document. 

To compare the low and high resolution reconstructions
in a quantitative manner, we down-sampled the 
DINR-512 reconstructed volumes from $512^3$ voxels to $128^3$ voxels.
This comparison also serves to demonstrate the superior reconstruction 
of lower resolution features using DINR-512.
We report the PSNR between each reconstruction 
and the associated ground-truth in Fig. \ref{fig:mpm_quant} (a). 
The PSNRs are averaged over $10$ volumes 
that are equally spaced in time. 
DINR-512 produces the highest PSNR for all datasets.
While the performance benefits of DINR-128 compared to PINR-128
are inconclusive, we observed that
the DINR-128 reconstructions contain fewer artifacts 
in the isosurface visualizations. 

We experimentally validated strong scaling of DINR 
using an IBM GPU compute cluster with 4 NVIDIA P100 GPUs per compute node \cite{LC_Lassen}. 
We report speedup for $30$ epochs of neural network 
optimization across different number of GPUs and compute nodes.
We measured the total run times as the number of nodes 
was varied from 4 (16 GPUs) to 32 (128 GPUs). 
If $D$ is the number of nodes, then we define speedup
as the ratio of the runtime using one node and $D$ nodes. 
As shown in Figure~\ref{fig:mpm_quant} (b), 
the scaling results demonstrate strong scalability 
of DINR due to effective distribution of the optimization 
across multiple nodes and GPUs. \textcolor{cgcol}{Additionally, we report the training time in minutes 
for DINR-512 in this figure.}
%The inference time to reconstruct $10\times128^3$ volume frames is approximately 11 minutes (66 seconds per 3D volume) using a single NVIDIA Titan Xp GPU}. 

\section{Discussion}\label{sec12}

DINR is a function approximator that expresses the
linear attenuation coefficient (LAC)
as a continuous function of the 
object coordinates.
Since the output of DINR is a continuous function in time-space,
it is inherently resistant to reconstruction of discontinuous 
artifacts such as streaks and noise unlike discrete voxel representations.
The SOTA approaches to experimental
4DCT reconstruction assume a static object for the duration 
of projection images used to reconstruct one volumetric time frame.
In this case, limited angle artifacts manifest 
as spurious streaks when the 
ratio of the total angular rotation for the projections 
to the number of time frames is low.
These spurious artifacts rotate 
with the changing angular direction of X-rays across view angles
and limit the temporal resolution of conventional 4DCT.
In contrast, DINR inherently assumes a dynamic object that
changes continuously over time and between subsequent projection images.
This allows DINR to achieve an order-of-magnitude higher
temporal resolution that approaches the time duration 
for acquiring a single projection image.
This capability of DINR is revolutionary since it allows 
4D imaging of very fast dynamic scenes or objects.
\textcolor{cgcol}{Since the output of INRs are 
continuous functions of the input coordinates,
they tend to produce outputs that vary smoothly with the coordinates.
Thus, INRs are inherently averse to reconstruction of spurious streak artifacts that typically plague 4DCT reconstructions.}

During each iteration of training, we only sample
a small subset of projection pixels to compute 
the updated values of the network parameters.
For instance, we sampled $0.002\%$ of the projection pixels
during each training iteration for the log-pile sample.
Then, we forward and back propagate through 
the DINR network at randomly sampled object coordinates 
along the ray paths connecting the sampled projection pixels to the source.
This approach to training is highly memory and compute efficient
since it does not require instantiating and computing the values at
voxels that span the complete space-time object coordinates.
We uniformly discretize the space-time coordinates during inference
and compute the reconstructed values at the discrete voxels
on a 4D time-space grid. 
While it may appear that the inference step is memory 
intensive, this memory is only allocated on hard disk drives 
that are typically very large in size (up to terabytes (TB)). 
In contrast, training utilizes GPU memory that is typically
very limited (up to few tens of gigabytes). 

While we can produce a reconstruction 
from the DINR at any arbitrary time instant,
we do not expect a higher time resolution than
the time gap between two adjacent projection images (or view angles).
Similarly, we do not expect a higher spatial resolution
than the size of each detector pixel 
that is back projected onto the object plane.
Thus, we only sample and voxelize at the time instants of each projection view
and at a spatial resolution equal to the detector pixel size
divided by the geometric magnification.
The geometric magnification is one for parallel-beam 
and is the ratio of the source-to-detector distance and
the source-to-object distance for cone-beam geometry. 

DINR is a compressed representation of the temporal and
spatial distribution of the LAC for the scanned object.
Information on this 4D reconstruction is encoded 
in the parameters of the DINR network. 
The number of network parameters is an order of magnitude less
than the number of discrete voxels across time-space.
For the log-pile and SiC samples, 
the numbers of voxels across time-space are approximately 
$3\times 10^{11}$ ($8.8$ TB) and $7\times 10^{10}$ ($2$ TB) respectively.
% for log-pile, size is 400*1024*1024*722*32/1024/1024/1024/1024
% for SiC, size is (765-437)*1024*1024*200*32/1024/1024/1024/1024
%However, the number of network parameters that reconstruct
%the values at these voxels is approximately $3\times 10^5$ ($10$ MB).
However, the number of trained network parameters used in DINR 
for the entire 4D reconstruction is approximately $3\times 10^5$ ($10$ MB).
% computed network parameter size as (256*256*5+256*5+256+1)*32/1024/1024

As supplementary materials,
we provide movies that show 4D reconstructions
of the experimental X-ray CT datasets.
Since 4D reconstructions show volumetric evolution
of samples over time, movies are an excellent
media for effective visualization of the reconstructions.
We show reconstructions at time steps
that correspond to all the acquisition times for the projection images.
Movie 1 shows the 4D high-resolution DINR reconstruction 
of the log-pile sample.
Movie 2 compares 4D reconstructions of the log-pile sample
using DINR and FDK.
Movie 3 shows the 4D high-resolution DINR reconstruction 
of the SiC sample.
Movie 4 compares 4D reconstructions of the SiC sample 
using DINR and FDK.
\textcolor{cgcol}{The FDK in 
Movies 2 and 4 use non-overlapping windows of
projection views.
Movie 5 compares 4D reconstructions of the log-pile sample
using DINR and FDK for overlapping windows of projection views.}

\section{Conclusion}

We formulated a novel approach called 
distributed implicit neural representation (DINR) for 
reconstruction of objects 
imaged using X-ray computed tomography.
DINR is primarily composed of a fully-connected neural network
that is trained to reconstruct the X-ray attenuation properties 
of the object at its output as a function of the time-space coordinates. 
For training the DINR network, we presented a novel stochastic 
distributed optimization algorithm with near-linear scaling capability 
as a function of the number of GPUs.
DINR is capable of producing terabyte sized 
4D reconstructions at high spatial and temporal resolutions.
Using 4DCT experimental data, we demonstrated that only DINR is
able to clearly resolve crack propagation in a SiC sample
and reconstruct the fine interleaving strands of a log-pile sample over time. 
We used the PSNR and SSIM performance metrics applied on
both simulated and experimental
data to quantitatively demonstrate the advantage of DINR when 
compared to SOTA 4DCT reconstruction methods.

\section{Acknowledgments}
LLNL-JRNL-860955. This work was funded by the Laboratory Directed Research and Development (LDRD) program at Lawrence Livermore National Laboratory (22-ERD-032). This work was performed under the auspices of the U.S. Department of Energy by Lawrence Livermore National Laboratory under contract DE-AC52-07NA27344. Lawrence Livermore National Security, LLC. 

\appendices

\section{Linear Projection}
\label{sec:linproj}

\begin{figure*}[!thb]
\begin{center}
\includegraphics[width=6in]{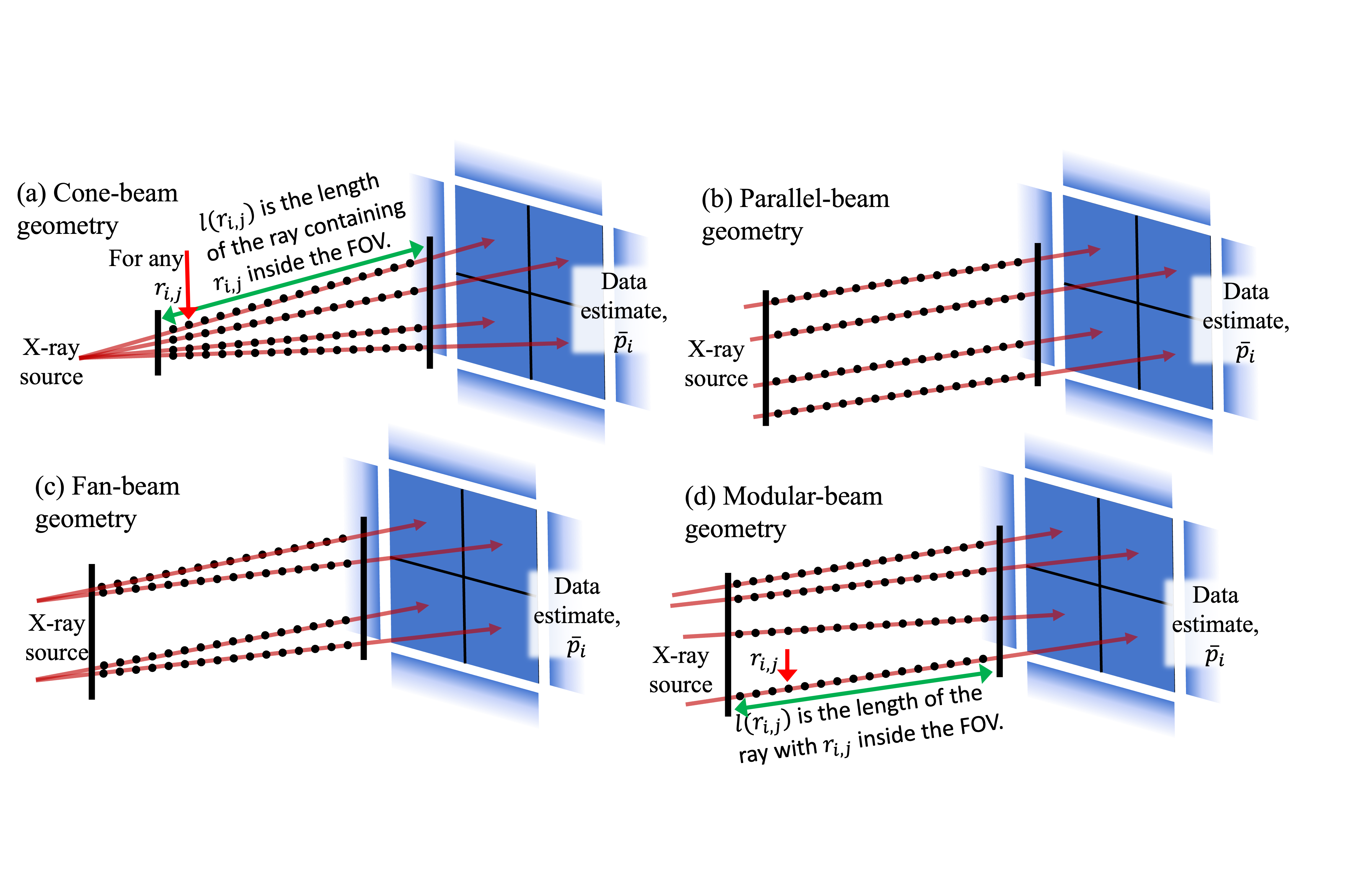}
\end{center}
\caption{\label{fig:mtdraysum}
$\bar{p}_i$ is the weighted sum of $\mathcal{M}\left(r_{i,j}; \gamma\right)$
at equi-spaced coordinate samples on multiple ray traces 
from the X-ray source to the detector pixel 
that measures the projection $p_i$. 
The contribution of $\mathcal{M}\left(r_{i,j}; \gamma\right)$
to $\bar{p}_i$ is scaled by the length of the portion 
of the ray inside the region-of-influence that contains $r_{i,j}$. 
We show ray traces for cone-beam, 
parallel-beam, fan-beam,
and modular-beam geometries in (a), (b), (c), and (d) respectively.
}
\end{figure*}

In this section, we present our approach to compute
$\bar{p}_i$ that is an estimate for the $i^{th}$
projection used in equation \eqref{eq:mainsqdist}.
To compute $\bar{p}_i$, we trace several rays 
from the X-ray source to the pixel that measures $p_i$
as shown in Fig. \ref{fig:mtdraysum}.
The end points of all the rays are equally spaced 
on a 2D grid over the surface of the $i^{th}$ pixel.
The 2D grid is obtained by equally sub-dividing the pixel 
into $D\times D$ number of sub-pixels and tracing rays to 
the centers of the sub-pixels. 
For a diverging cone-beam (Fig. \ref{fig:mtdraysum} (a)), the rays
originate from a single point source.
For parallel-beam (Fig. \ref{fig:mtdraysum} (b)), the rays
are mutually parallel.
We also show ray tracings for fan-beam
and modular-beam geometries in Fig. \ref{fig:mtdraysum} (c)
and Fig. \ref{fig:mtdraysum} (d) respectively.
However, we do not investigate reconstructions from 
fan-beam and modular-beam geometries in this paper.
Modular-beam is a generic specification
for defining the direction of X-ray propagation
that applies to any system geometry. 
To define the geometry, it uses known locations for the X-ray sources
and detector panels that can be placed at arbitrary coordinates in 3D space.

Along each ray in Fig. \ref{fig:mtdraysum}, 
we produce equi-spaced samples of coordinates
that are within the cylindrical field-of-view
for the object.
The spacing between adjacent coordinate samples
along each ray is approximately $\Delta/D$, 
where $\Delta$ is the projection pixel size that is back-projected onto the object plane and $D$ is the up-sampling factor (chosen as $D=2$ in this paper).
We denote each coordinate sample as $r_{i,j}$.
Thus, the estimate $\bar{p}_i$ is given by,
\begin{equation}
    \bar{p}_i = \frac{1}{\vert\Phi_i\vert}\sum_{j\in\Phi_i} w(r_{i,j})\mathcal{M}(r_{i,j}; \gamma).
\end{equation}
Here, the weight term is expressed as 
$w(r_{i,j})=\mu_0 l(r_{i,j})$ and is defined in
equation \eqref{eq:weightfactors}. 
The purpose of $\mu_0$ is to produce LAC reconstructions
in correct units since the random initialization 
used for the neural network layers
serve to produce normalized outputs that are unit less. 

During back-propagation, each process computes the 
gradient of $L\left(\Omega_k; \gamma\right)$ 
with respect to the vector of network parameters, $\gamma$.
This gradient is another vector that is denoted by 
$\nabla_{\gamma}L\left(\Omega_k; \gamma\right)$
whose $l^{th}$ element is the partial derivative 
$\partial L\left(\Omega_k; \gamma\right)/\partial \gamma_l$
that is given by,
\begin{multline}
    \label{eq:partiald}
    \frac{\partial L\left(\Omega_k; \gamma\right)}{\partial \gamma_l}
    = \frac{1}{\vert\Omega_k\vert}\sum_{i\in\Omega_k}
    2 \left(p_i-\frac{1}{\vert\Phi_i\vert}\sum_{j\in\Phi_i}  w(r_{i,j})\mathcal{M}(r_{i,j}; \gamma)\right)\\
    \left(-\frac{1}{\vert\Phi_i\vert}\sum_{j\in\Phi_i} w(r_{i,j})\frac{\partial \mathcal{M}(r_{i,j}; \gamma)}{\partial \gamma_l}\right).
\end{multline}
Note that equation \eqref{eq:partiald} is not
explicitly implemented in code, but is instead automatically
computed using algorithmic differentiation in PyTorch.

\section{Network Architecture}
\label{sec:netarch}

\begin{figure*}[h]
\centering
\includegraphics[width=0.99\linewidth]{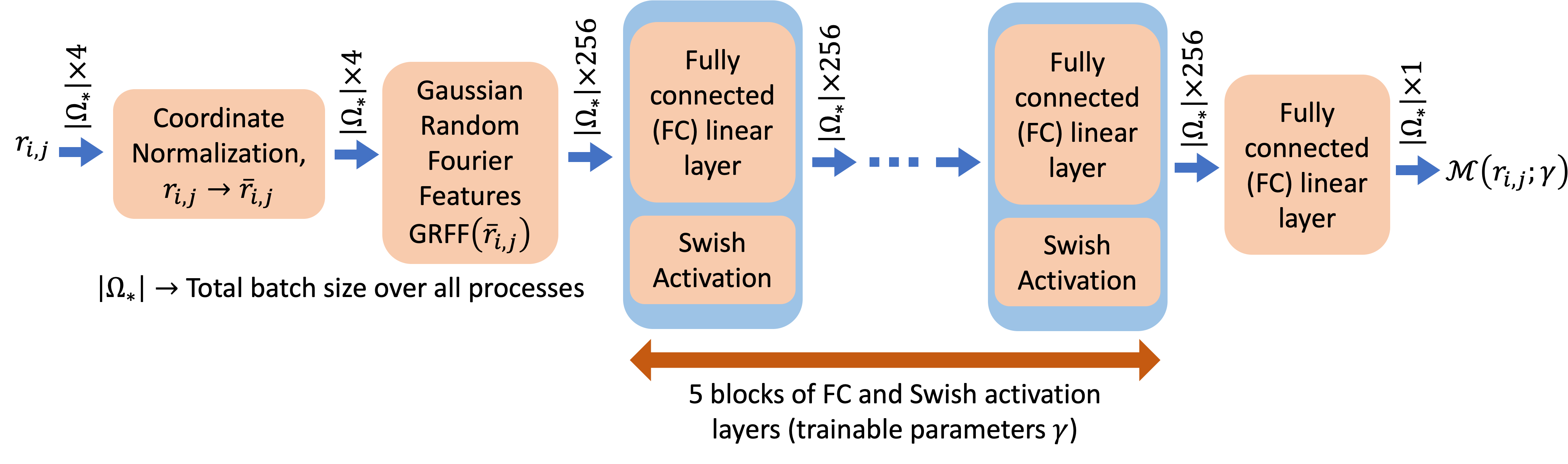}
\caption{Detailed Network architecture for DINR.}
\label{fig:nnarch_detail}
\end{figure*}

The DINR network, $\mathcal{M}(r_{i,j}; \gamma)$, 
is the concatenation of the following blocks of layers,
\begin{enumerate}
    \item Normalize $r_{i,j}=\left(t_i, z_{i,j}, y_{i,j}, x_{i,j}\right)$ such that each normalized coordinate in $r_{i,j}$ 
    is between $-1$ and $1$. 
    Since our object is assumed to lie inside the cylindrical field-of-view
    shown in Fig. \ref{fig:mainfig}, we normalize using the minimum
    and maximum coordinate values of the cylindrical boundary.
    Let $\bar{r}_{i,j}$ denote the normalized coordinates.
    \item Gaussian random Fourier feature (GRFF) encoding layer
    \cite{Fourier_INR_NEURIPS2020}
    to map the coordinates $\bar{r}_{i,j}$ to GRFF features as,
    \begin{equation}
        \text{GRFF}(\bar{r}_{i,j}) = 
        \begin{bmatrix}
            \cos\left(2\pi B \bar{r}^T_{i,j}\right) \\
            \sin\left(2\pi B \bar{r}^T_{i,j}\right)
        \end{bmatrix}^T
    \end{equation}
    where $T$ denotes transpose,  $\bar{r}^{T}_{i,j}$ is a column vector, 
    $B$ is a $C \times 4$ matrix of random numbers.
    Each element along the $1^{st}$ column of $B$ is 
    sampled from a Gaussian distribution with zero mean 
    and standard deviation of $\sigma_t$.
    Elements along the $2^{nd}$, $3^{rd}$, and $4^{th}$ columns of $B$
    are sampled from a Gaussian distribution with zero mean
    and standard deviation of $\sigma_s$.
    The elements of $B$ are not modified during training of 
    $\mathcal{M}(r_{i,j}; \gamma)$.
    An increasing value for $\sigma_s$ leads to
    increasing sharpness and noise of the DINR reconstruction 
    along the three spatial dimensions and vice versa.
    Similarly, increasing $\sigma_t$ leads to increasing 
    sharpness and noise of the DINR reconstruction along the time dimension
    and vice versa. 
    Thus, the value of $\sigma_s$ and $\sigma_t$ are important parameters
    that trade off sharpness with noise and spurious artifacts.
    \item The GRFF outputs of length $2C$ are then input to a sequence
    of $L$ fully connected (FC) layers.
    Each FC layer consists of a linear layer 
    followed by a 
    Swish activation function \cite{swish_ramachandran2017}.
    The number of input and output channels
    for each FC layer is $2C$.
    \item At the end, we have a fully connected linear layer to map the $2C$ number of input 
    channels to a single scalar value that is the object reconstruction
    at coordinate $r_{i,j}$. We multiply this value by $\mu_0$ 
    to convert it to LAC reconstruction units.
    We do not use an activation function for this last FC layer.
\end{enumerate}
\textcolor{cgcol}{A schematic of the network architecture is
shown in Fig. \ref{fig:nnarch_detail}.
The training loop is shown in Fig. \ref{fig:nnarch_train}
of the supplementary document.}
\textcolor{cgcol}{The choice of the number of channels 
and layers is a trade off between sufficient representational
capacity and sparsity of representation.
If the loss function during training does not reduce by an order
of magnitude during training, it may imply the need to 
increase representational capacity by increasing the number
of layers and channels for achieving good reconstruction quality.
However, the magnitude of loss reduction is also determined by
the noise in the measurements. For high noise data, the 
reduction in loss may not be substantial, but may still achieve
excellent reconstructions. Alternatively, a large number of layers
and channels may imply more network parameters with 
insufficient sparsity. Insufficient sparsity may lead DINR 
to reconstruct undesired artifacts such as streaks.}

\section{Training Parameters}
\label{sec:trainpars}
For the experimental data reconstruction in Fig. \ref{fig:deben202303}, 
we set $\sigma_s=0.5$ and $\sigma_t=0.1$, 
\textcolor{cgcol}{which
is a also a suitable default that worked well for
all our tested datasets. It led to excessive smoothing for other datasets, but is nevertheless a good default.}
For the experimental data reconstruction in Fig. \ref{fig:deben202310}, 
we set $\sigma_s=5.0$ and $\sigma_t=0.1$.
If $\sigma_t$ is progressively increased beyond a certain threshold,
we notice artifacts where the spatial features of the object 
seem to also rotate between subsequent time indices.
To get a stable reconstruction of the object,
we determined that $\sigma_t$ needs to be sufficiently small.
For our experimental data reconstructions, a single value 
of $\sigma_t=0.1$ was sufficient to get a stable reconstruction.
We recommend a coarse tuning of both $\sigma_s$ and $\sigma_t$ 
 by factors of $10$ starting from initial values of $1$
 before further fine tuning.
Our goal was only to determine workable parameter values
that produced a minimum desired visual reconstruction quality. 
It is difficult to estimate the optimal values for $\sigma_s$ and $\sigma_t$
since they are application dependent.

The number of channels for the FC layers is set as $2C=256$.
The number of FC layers with the non-linear Swish activation
function is $L=5$.
The batch size per compute process is $\vert\Omega_k\vert=48$ 
in equation \eqref{eq:localoptfunc}.
We used $128$ number of compute processes to train the DINR
used for Fig. \ref{fig:deben202303} and \ref{fig:deben202310}.
The $128$ processes are distributed across $32$ HPC nodes each with $4$
Nvidia Tesla V100 GPUs such that each process runs on one GPU.
The learning rate is $0.001$ at the beginning of the training loop.
We use learning rate decay to progressively reduce the learning
rate by a multiplicative factor of $0.95$ after each epoch.


\bibliographystyle{IEEEtran}
\bibliography{IEEEabrv, paper}

%\newpage

 

% \end{IEEEbiography}



\newpage
\section{Supplementary Document}
\subsection{Preprocessing of Experimental Data}\label{secEDPre}
\begin{figure*}[!thb]
\begin{center}
\begin{tabular}{cc}
     \includegraphics[width=2.5in]{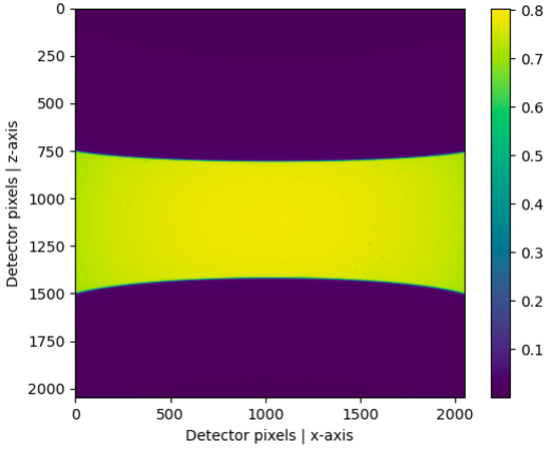} & 
     \includegraphics[width=3in]{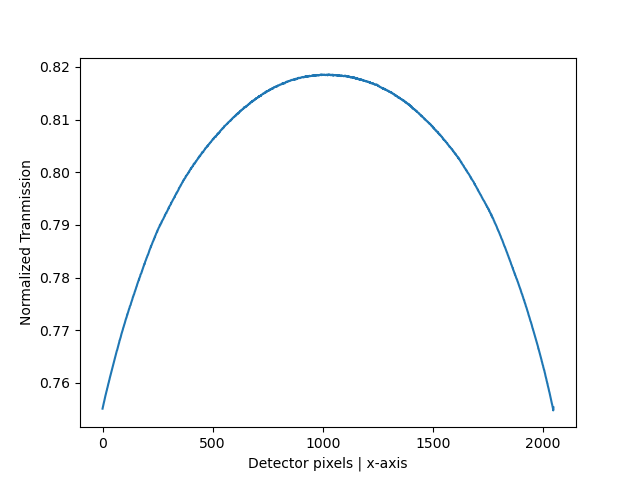} \\
     (a) & (b) \\
\end{tabular}
\end{center}
\caption{\label{fig:supp_deben_preproc}
(a) shows a radiograph (X-ray transmission image) of the Deben stage without the object.
The dark regions at the top/bottom are the anvils and the center bright region
is an envelope that surrounds the object.
(b) shows the average of several rows of pixels from the central bright region in (a). 
(b) is used to calibrate the radiographs from the dynamic CT scans. 
}
\end{figure*}

Fig. \ref{fig:supp_deben_preproc} (a) shows the normalized transmission of X-rays through only the Deben stage in the absence of the object. The near-zero transmission values at the top and the bottom of the image are a result of the complete attenuation of X-rays by the hardened steel anvils (platens) of the Deben stage. 
The regions of the anvils do not intersect the X-ray paths through the object and are excluded from the reconstruction’s field of view. The central region corresponds to the transmission of X-rays through the cylindrical carbon fiber envelope surrounding the Deben stage. The transmission value is lowest at the left/right corners and increases monotonically towards the image center. 
Fig. \ref{fig:supp_deben_preproc} (b) is the average of several rows of pixels from the non-zero transmission region of the Deben stage’s cylindrical envelope. We average several rows near the center of the envelope. Once the dynamic X-ray CT scans are acquired for the object inside the Deben stage, we divide each row of each radiograph in the CT scan by the curve in  Fig. \ref{fig:supp_deben_preproc} (b). 
We used this calibration procedure for the experimental CT datasets in Fig. \ref{fig:debendata}.
This calibration procedure corrects the CT projection images such that it only includes the attenuation by the object.

\subsection{Training Loop}
\begin{figure*}[h]
\centering
\includegraphics[width=0.99\linewidth]{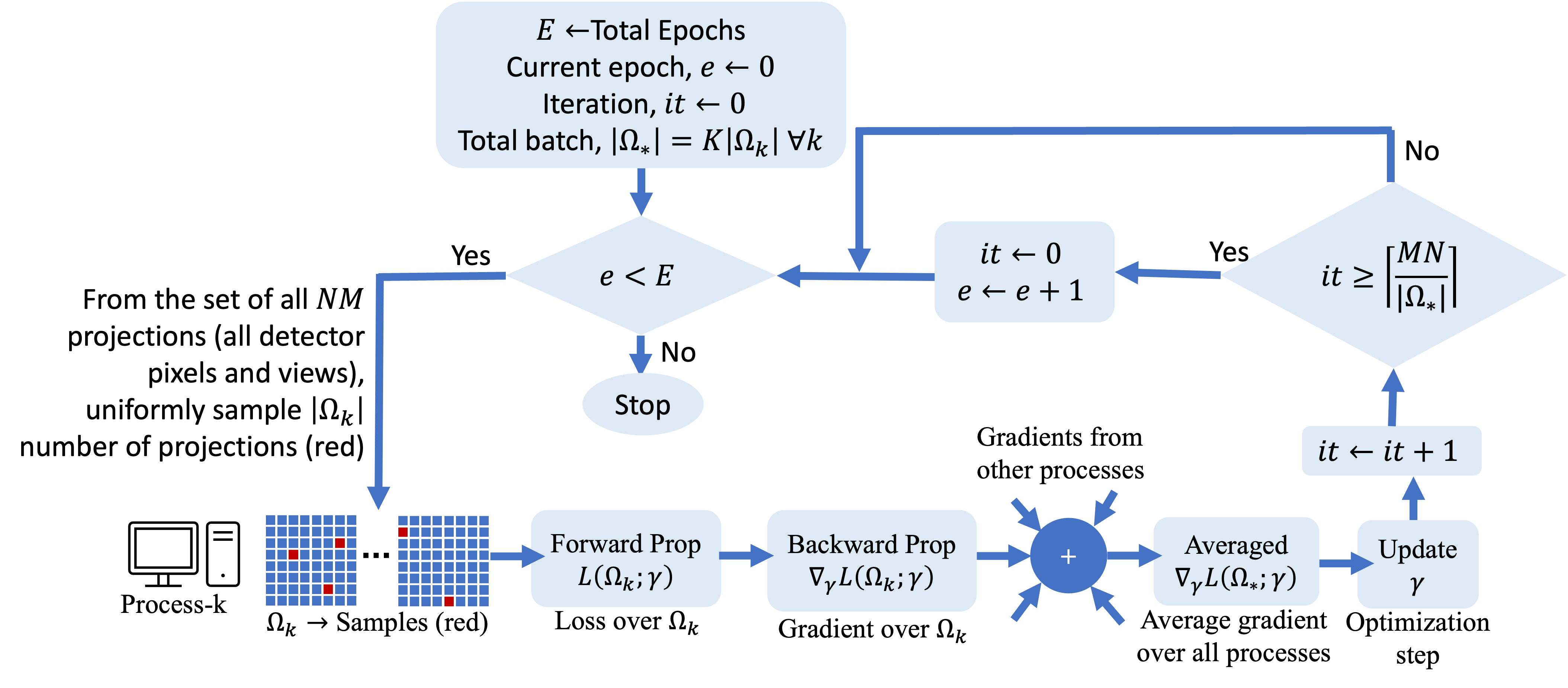}
\caption{Training loop for DINR.}
\label{fig:nnarch_train}
\end{figure*}

The overall training loop for DINR
is shown in Fig.  \ref{fig:nnarch_train} (b).

\subsection{Ablation Studies}\label{secA1}

In this section, we present ablation studies using simulated 4DCT data by performing several experiments to provide a more comprehensive understanding of our DINR reconstruction method.

\begin{figure*}[h]
\centering

\begin{tabular}[b]{c}
    \includegraphics[width=.6\linewidth]{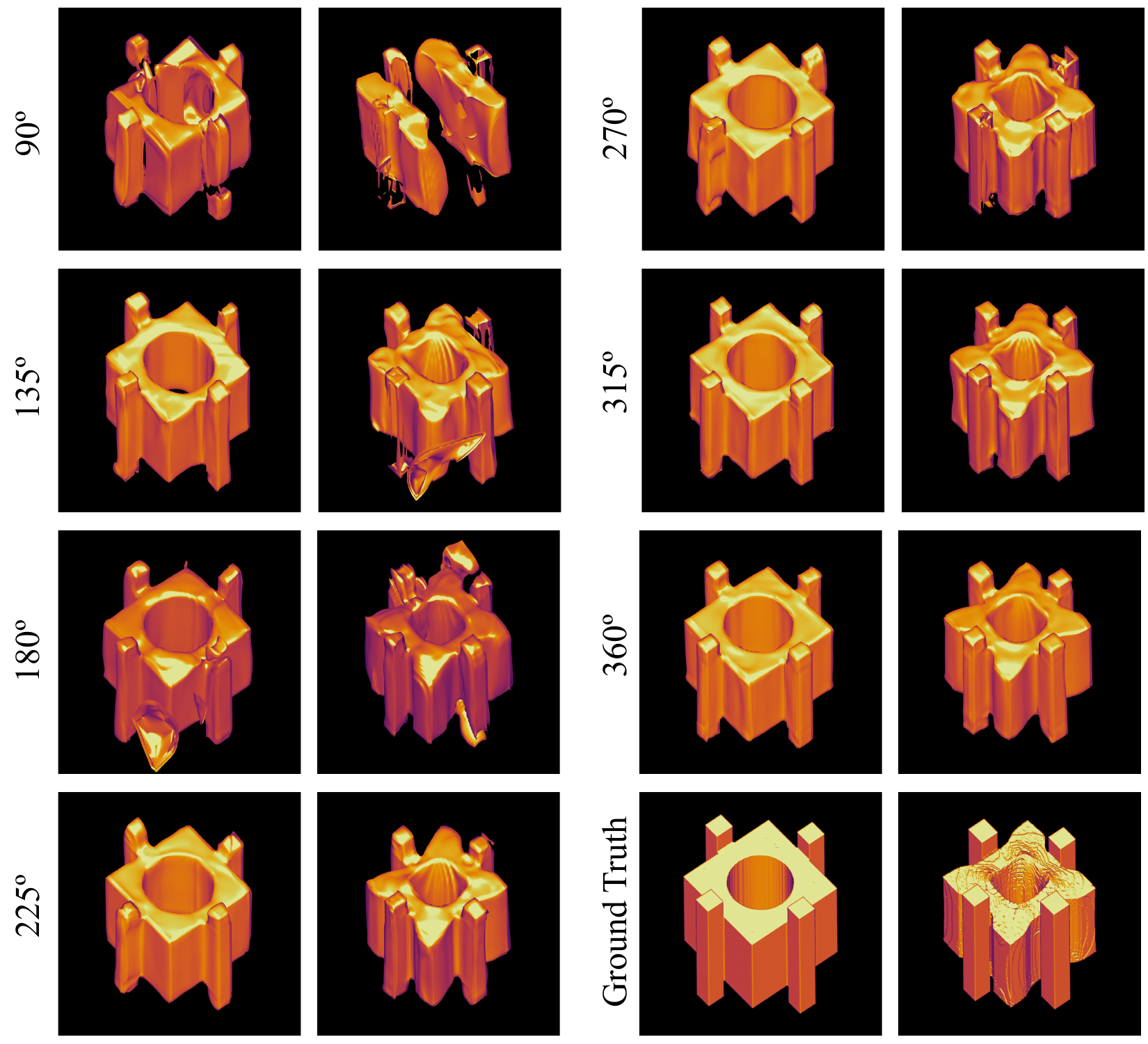} \\
    (a) ISO for first/last time frames.
    \label{fig:ablation_arange_iso}
\end{tabular}
\hfill
\begin{tabular}[b]{c}
    \includegraphics[width=.38\linewidth]{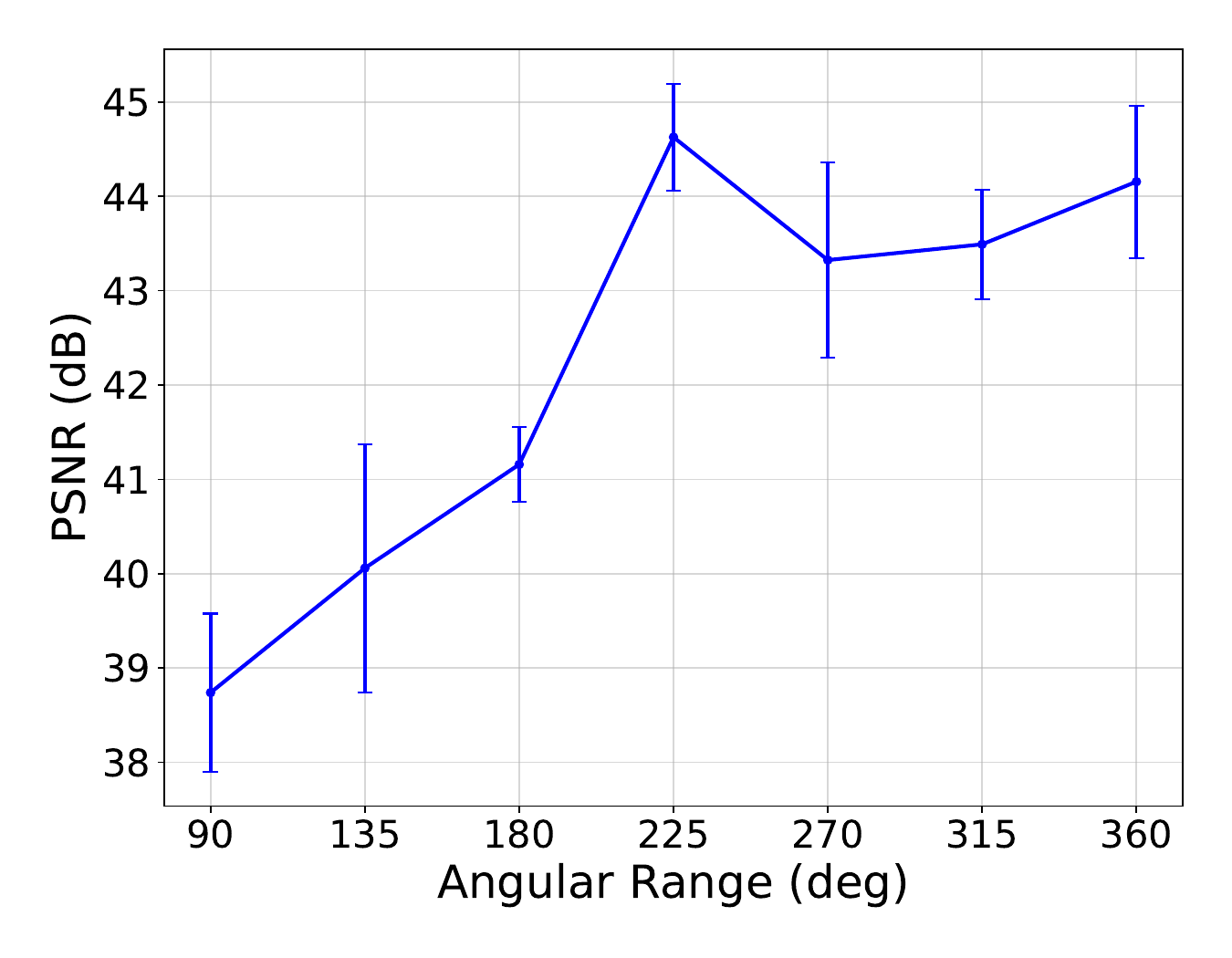} \\
    (b) PSNR vs. angular range.
    \label{fig:ablation_arange_psnr}
\end{tabular}

\caption{Effect of the rotation angular ranges on the reconstruction quality using our proposed DINR. (a) Isosurface 3D rendering of the first and last frame reconstructions for the rotation ranges shown in the row labels. (b) PSNRs of the reconstruction as a function of the different angular ranges.}
\label{fig:ablation_arange}
\end{figure*}

\subsubsection*{Effect of Angular Ranges}

When the rotation angular range is limited, 
the projection data is incompletely sampled.
This results in a poor reconstruction that is
plagued by significant artifacts. 
In the case of MPM-simulated datasets 
where rapid deformation occurs, 
it is considered severely limited-angle 4DCT 
even for a total rotation range of $360^{\circ}$. 
In this ablation study, we compared our DINR reconstructions 
for different angular ranges of the projection data 
to observe the reconstruction performance. 
As shown in Figure~\ref{fig:ablation_arange}, 
the reconstruction quality remains satisfactory 
even up to an angular range of $225^{\circ}$.

\begin{figure*}[h]
\centering

\begin{tabular}[b]{c}
    \includegraphics[width=.6\linewidth]{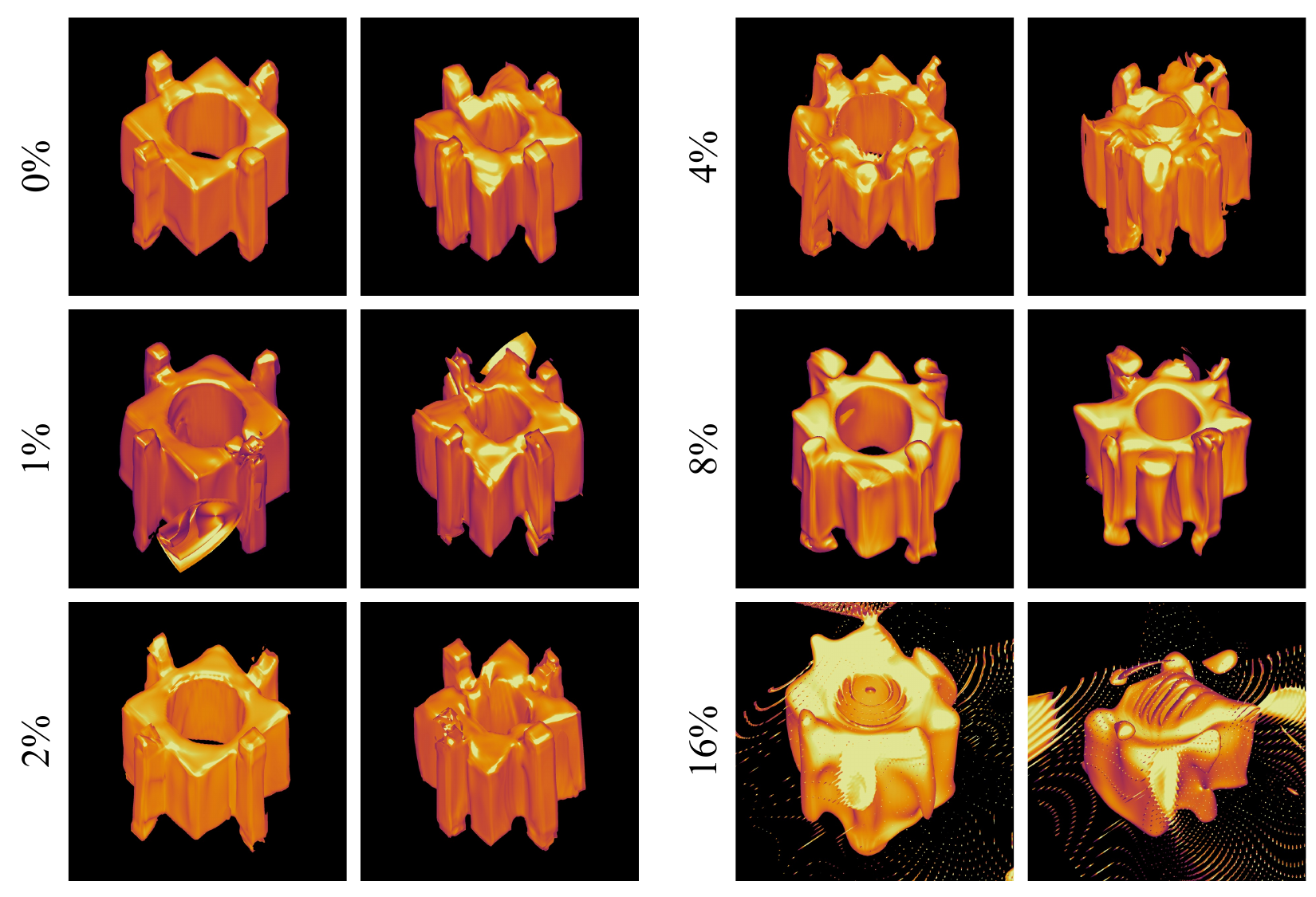} \\
    (a) ISO for first/last time frames.
    \label{fig:ablation_noise_iso}
\end{tabular}
\hfill
\begin{tabular}[b]{c}
    \includegraphics[width=.38\linewidth]{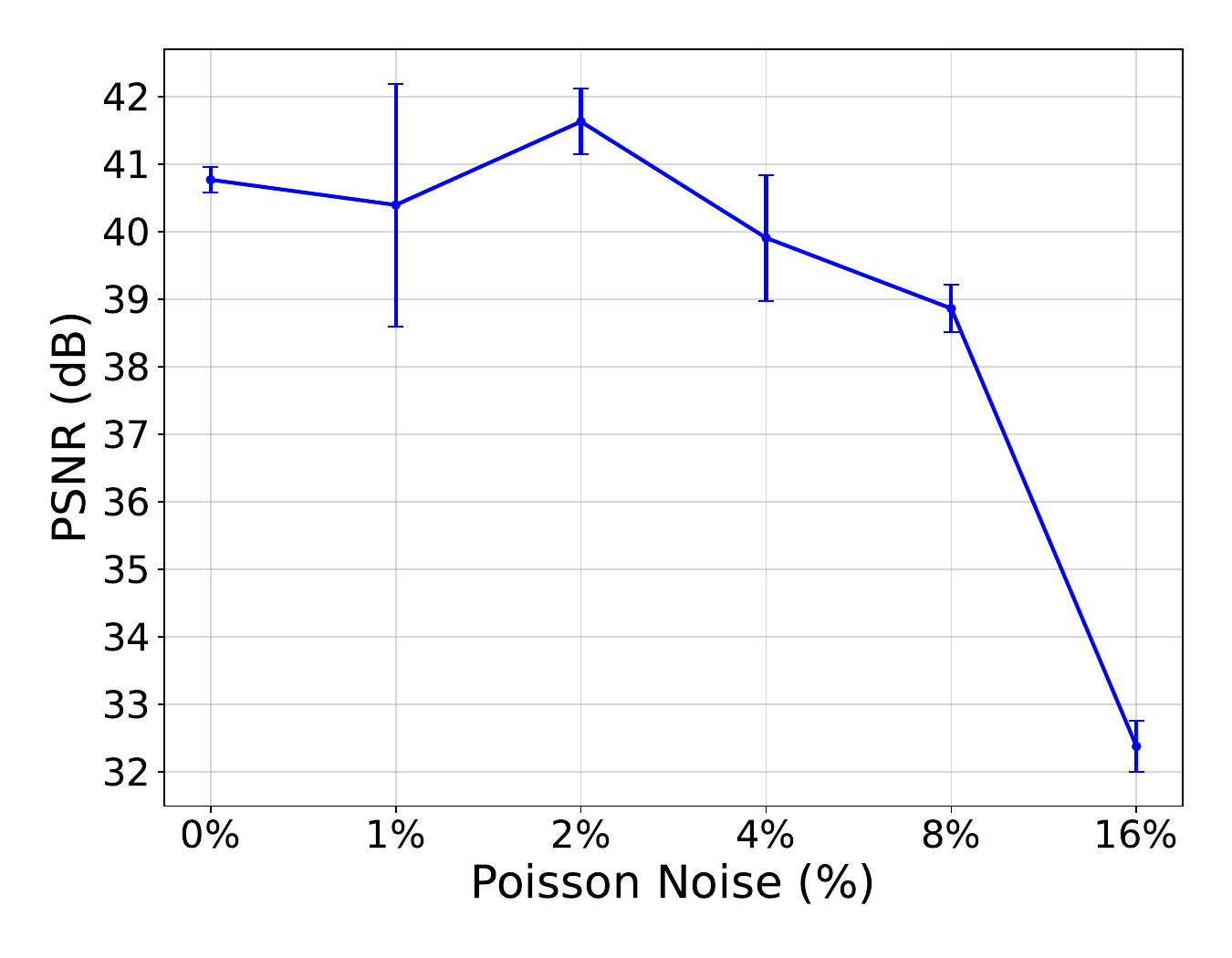} \\
    (b) PSNR vs. Noise.
    \label{fig:ablation_noise_psnr}
\end{tabular}

\caption{Effect of noise in projection data on the quality of reconstruction using our proposed DINR. (a) Isosurfaces of the reconstructed first and last frames for different noise levels specified in the row labels. (b) PSNRs of the reconstruction as a function of the noise percentages.}
\label{fig:ablation_noise} 
\end{figure*}

\subsubsection*{Effect of Noise in Projection Data}
For the reconstruction results in the main paper, 
we simulated CT 
projection data of the MPM-simulated datasets 
with a default noise level of 0.1\%. 
We add Poisson noise in the X-ray transmission space
(negative exponential of projections).
In this ablation study, we simulated 
different noise levels for the projection data 
to investigate the robustness of our DINR method. 
Noise was added to the projection data 
at 6 different noise levels: 0, 1, 2, 4, 8, 16\%, 
as shown in Fig~\ref{fig:ablation_noise}. 
We observe that the reconstruction quality 
remains satisfactory up to a noise level of 2\%.

\begin{figure*}[h]
\centering
\includegraphics[width=\linewidth]{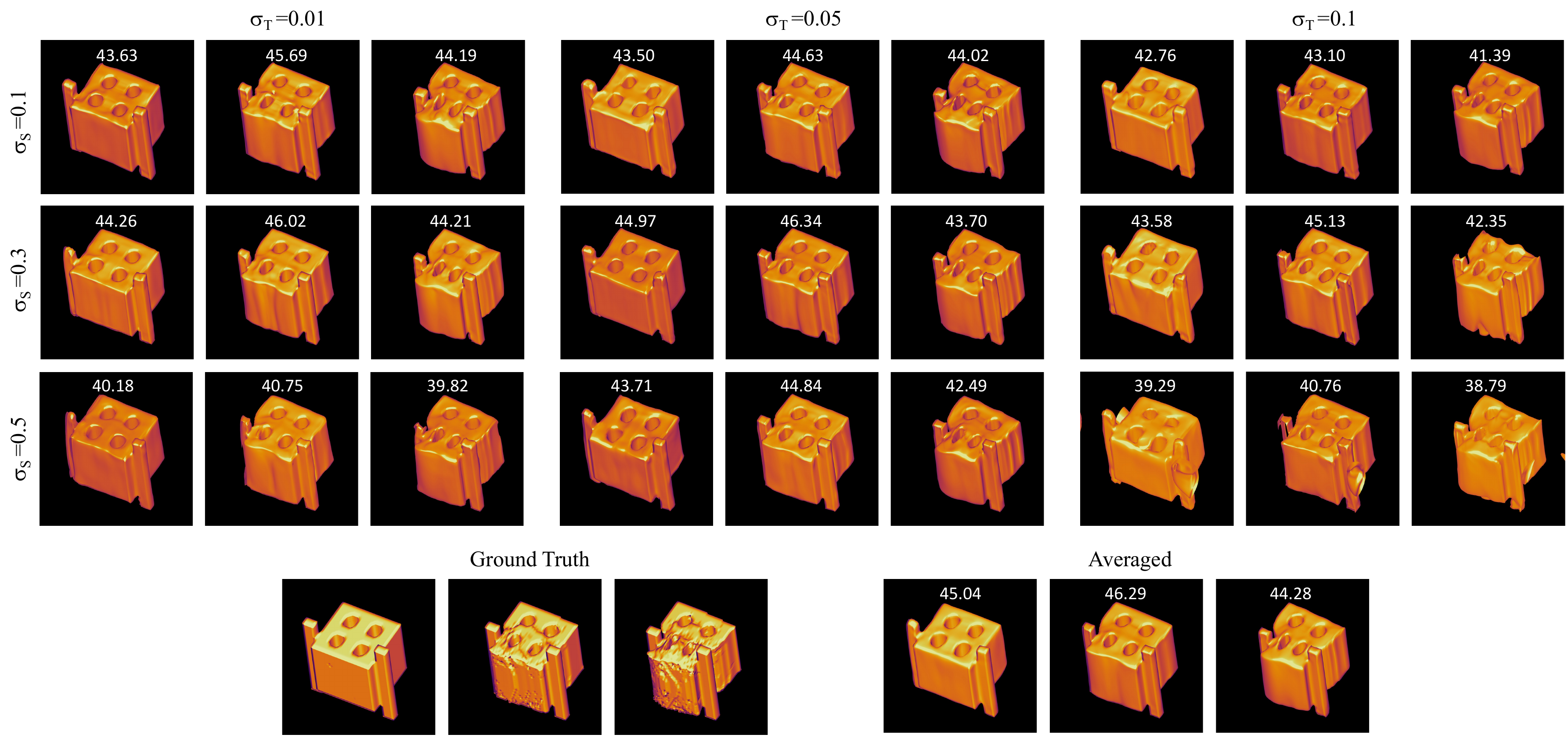}
\caption{Effect of different combinations of the GRFF parameters, $\sigma_s$ and $\sigma_t$, on the quality of reconstruction using DINR.}
\label{fig:ablation_sigma}
\end{figure*}

\subsubsection*{Effect of GRFF Parameters}

$\sigma_t$ and $\sigma_s$ are application-specific 
bandwidth parameters used by the 
Gaussian Random Fourier Features (GRFF) of DINR
to control the temporal and spatial smoothness respectively
for the 4D reconstruction.
These parameters function as regularization parameters. 
A careful selection of $\sigma_s$ and $\sigma_t$ parameters 
may be required to obtain an optimal reconstruction quality.
While an optimal setting for $\sigma_s$ and $\sigma_t$ 
varies depending on the scene, we observed that  
$\sigma_t$ ranging from $0.01$ to $0.1$ 
and $\sigma_s$ ranging from $0.1$ to $0.5$ yield 
reasonably good reconstruction quality for the MPM-simulated datasets. 
Figure~\ref{fig:ablation_sigma} shows 4D reconstructions
using DINR for different combinations of $\sigma_t$ and $\sigma_s$.

\subsection{More Reconstruction Results}\label{secB1}

\begin{figure*}
\centering

\subfloat[S03\_001\label{fig:supp_mpm1_a}]{
       \includegraphics[width=0.9\linewidth]{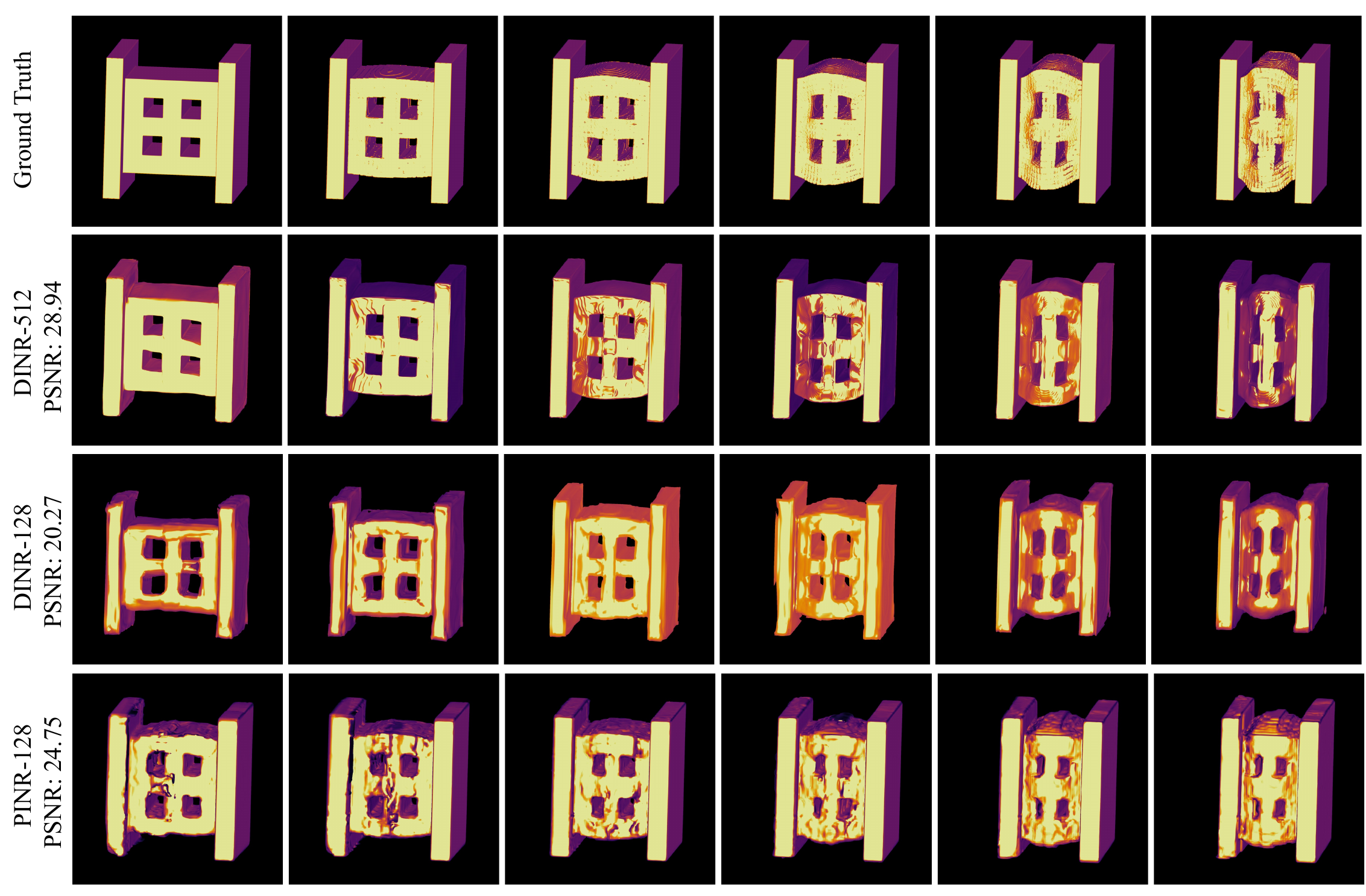}}

\subfloat[S03\_008\label{fig:supp_mpm1_b}]{
       \includegraphics[width=0.9\linewidth]{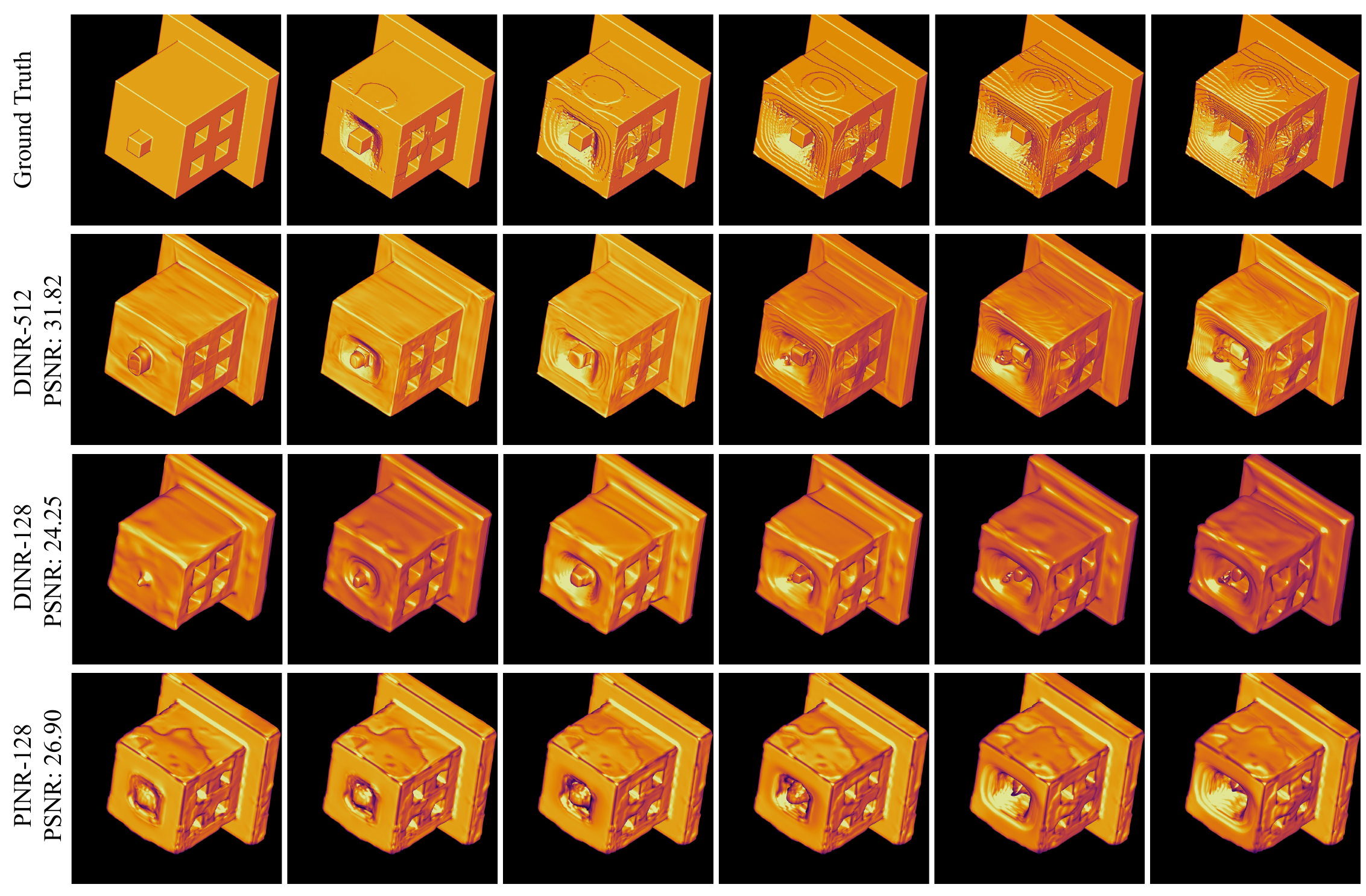}}

\caption{4D reconstruction of the MPM-simulated dataset: S03\_001 (a) and S03\_008 (b). From \textit{left} to \textit{right}, we show frames at $T_{0}, T_{20}, T_{40}, T_{60}, T_{80},$ and $T_{90}$ respectively.}
\label{fig:supp_mpm1} 
\end{figure*}

\begin{figure*}
\centering

\subfloat[S04\_009\label{fig:supp_mpm2_a}]{
       \includegraphics[width=0.9\linewidth]{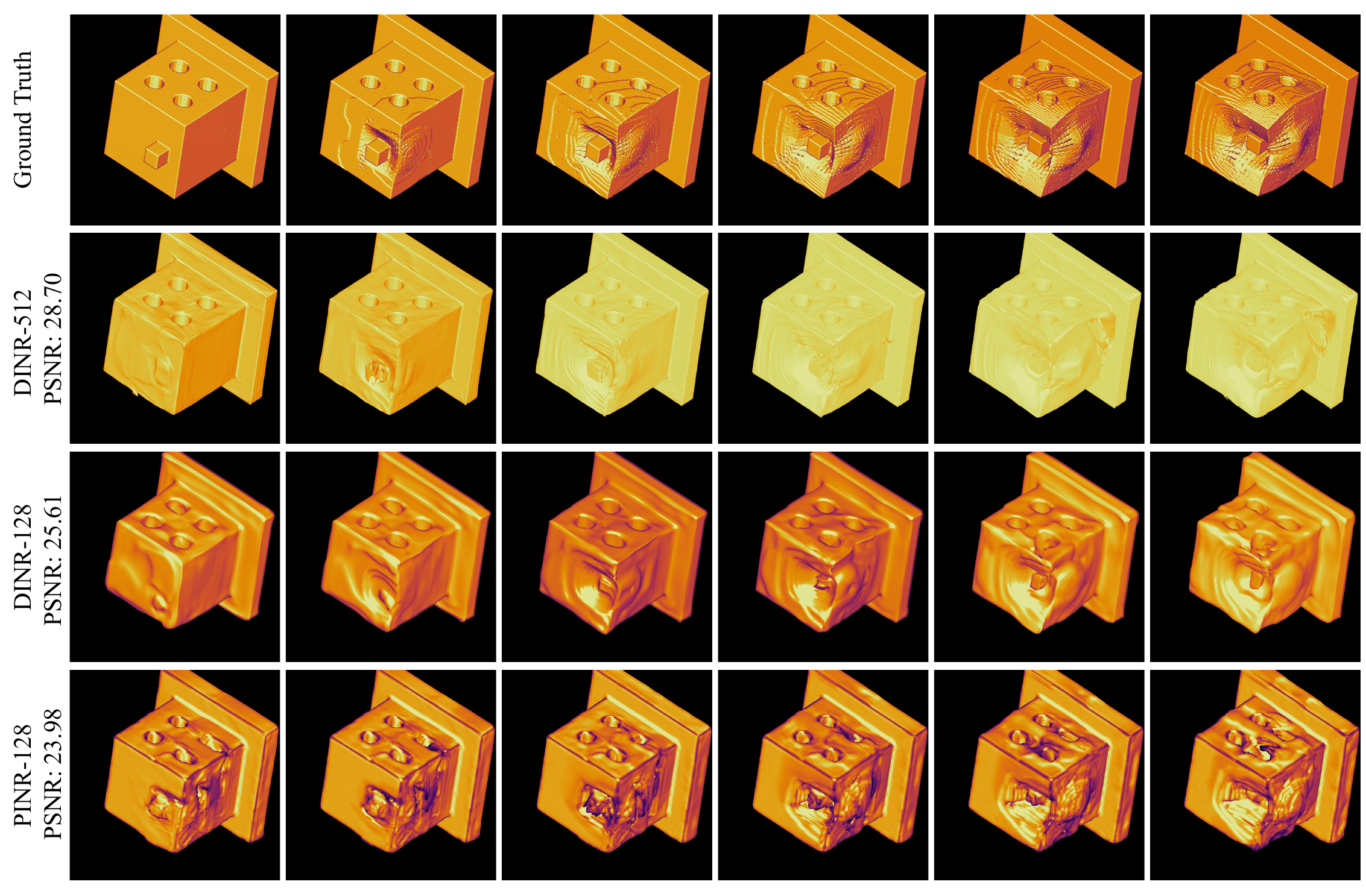}}

\subfloat[S04\_018\label{fig:supp_mpm2_b}]{
       \includegraphics[width=0.9\linewidth]{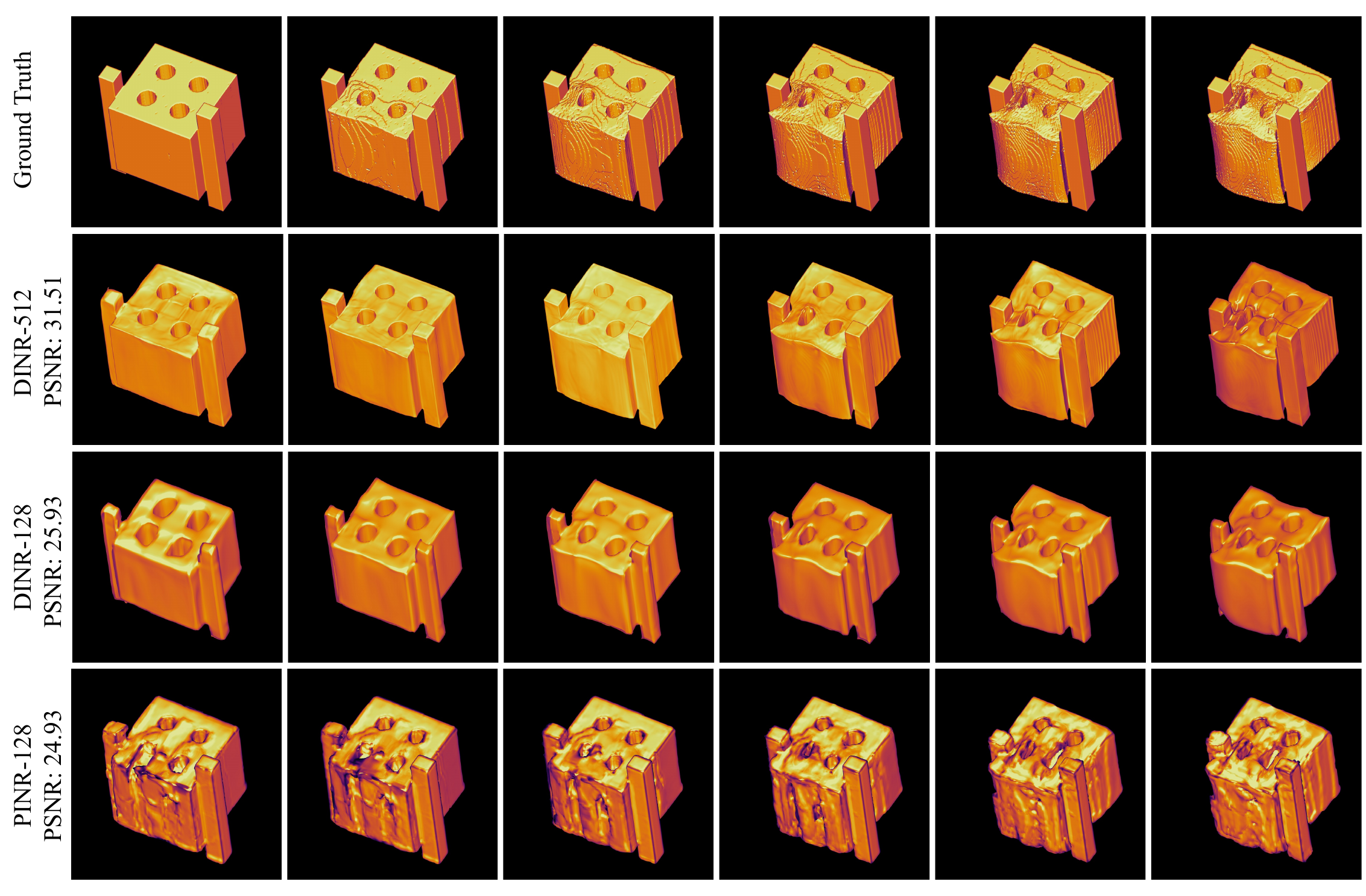}}

\caption{4D reconstruction of the MPM-simulated dataset: S04\_009 (a) and S04\_018 (b). From \textit{left} to \textit{right}, we show frames at $T_{0}, T_{20}, T_{40}, T_{60}, T_{80},$ and $T_{90}$ respectively.}
\label{fig:supp_mpm2} 
\end{figure*}

\begin{figure*}
\centering

\subfloat[S05\_700\label{fig:supp_mpm3_a}]{
       \includegraphics[width=0.9\linewidth]{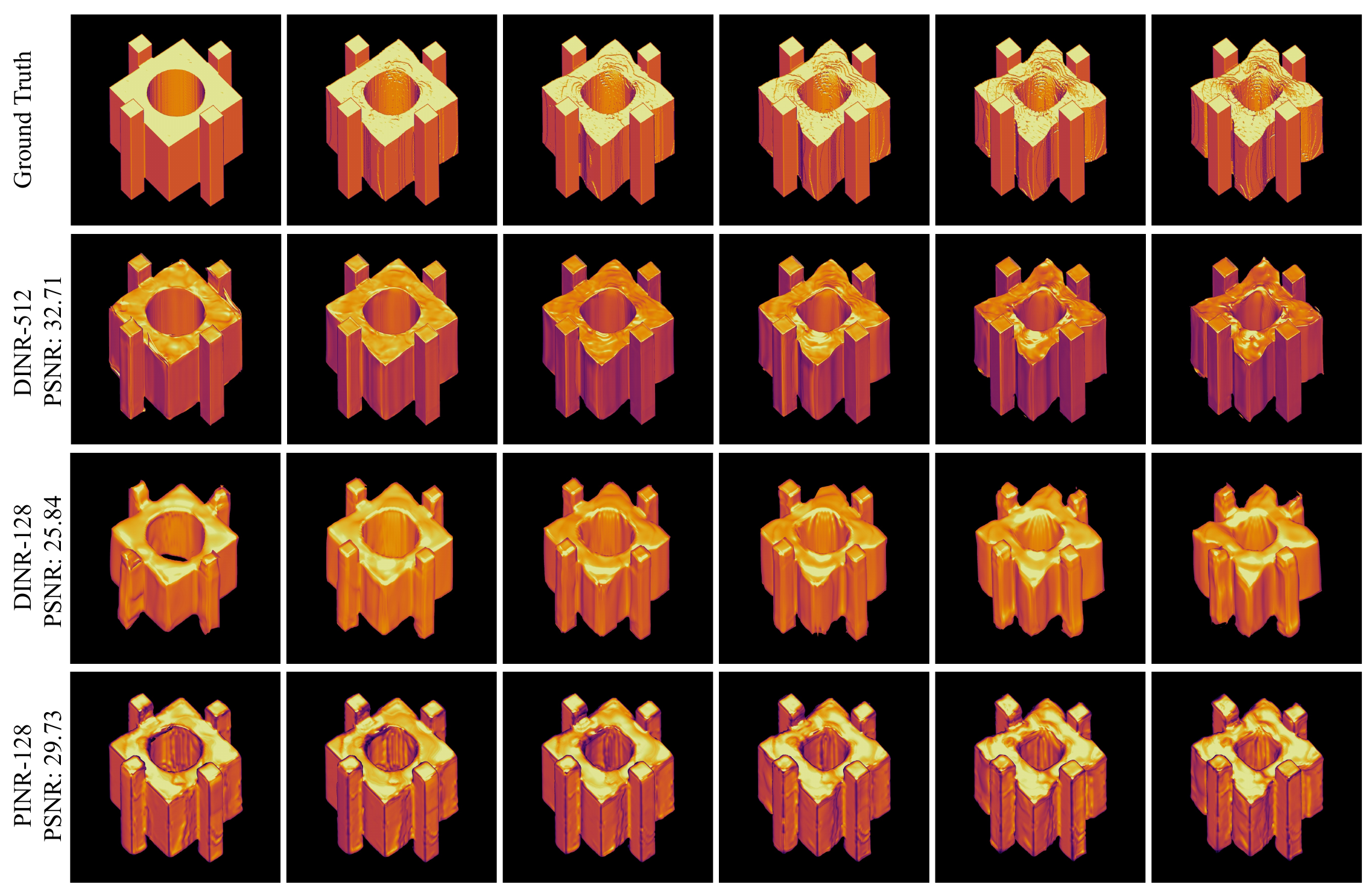}}

\subfloat[S08\_005\label{fig:supp_mpm3_b}]{
       \includegraphics[width=0.9\linewidth]{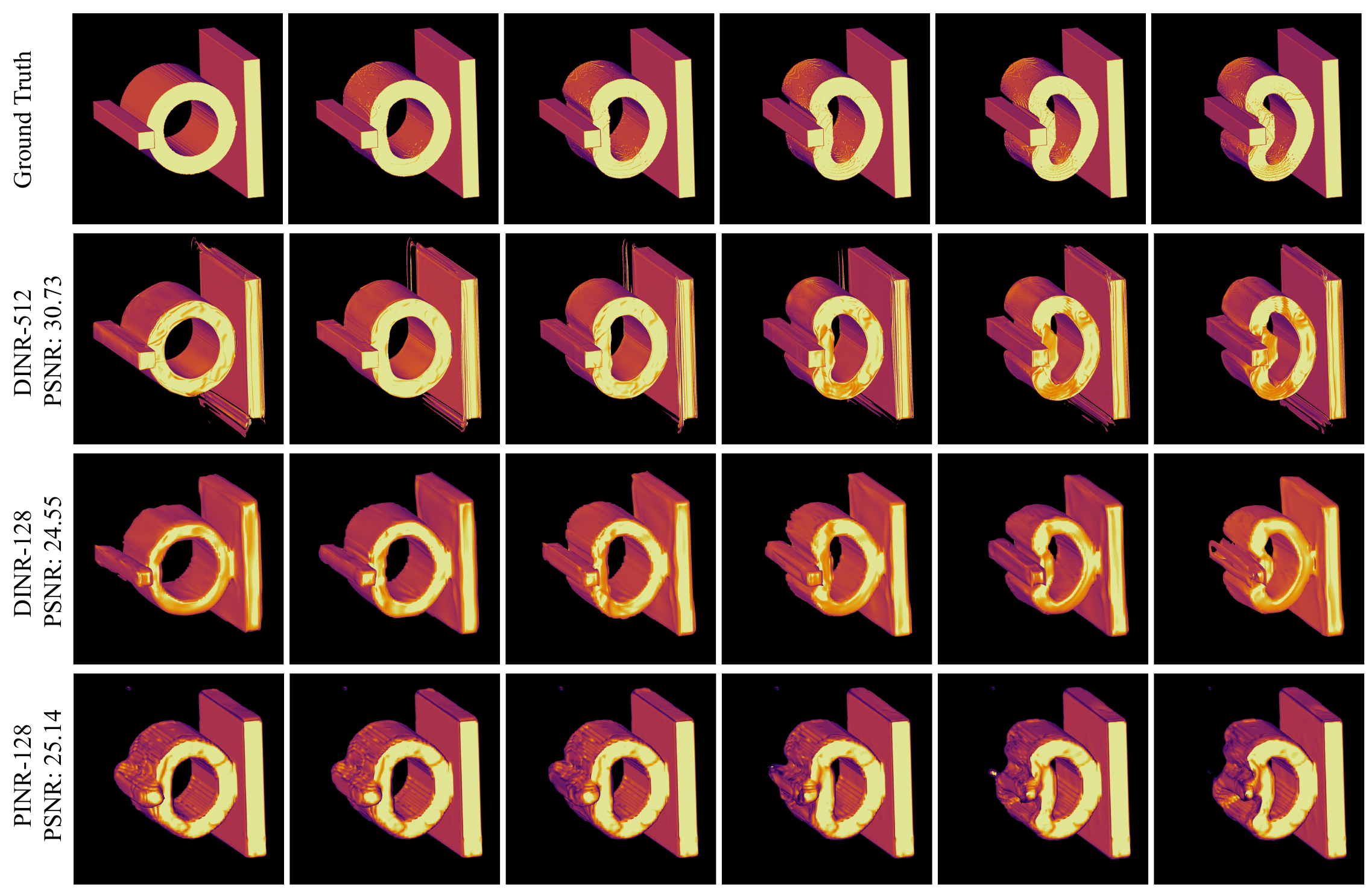}}

\caption{4D reconstruction of the MPM-simulated dataset: S05\_700 (a) and S08\_005 (b). From \textit{left} to \textit{right}, we show frames at $T_{0}, T_{20}, T_{40}, T_{60}, T_{80},$ and $T_{90}$ respectively.}
\label{fig:supp_mpm3} 
\end{figure*}

We show more reconstruction results for the MPM-simulated datasets that we did not include in the main paper 
in Fig. \ref{fig:supp_mpm1}, Fig. \ref{fig:supp_mpm2},
and Fig. \ref{fig:supp_mpm3} respectively.

\begin{figure*}[h]
\centering
    \includegraphics[width=0.4\linewidth]{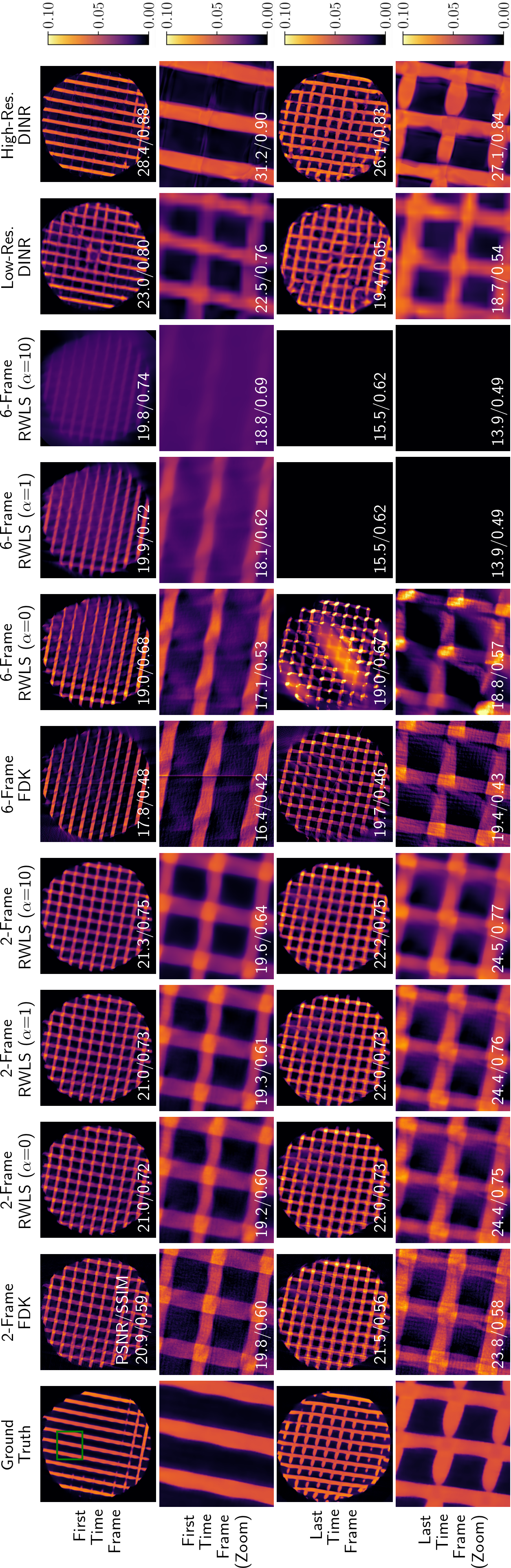}
\caption{\textcolor{cgcol}{Comparison between FDK, RWLS, and DINR (ours) for the log-pile experimental 4DCT dataset. This figure adds RWLS 
to Fig. \ref{fig:deben202303}. RWLS is regularized weighted least squares algorithm for CT reconstruction that uses
total variation regularization across 3D space \cite{LEAPCT}.}}
\label{fig:log-pile-fdk-rwls-dinr}
\end{figure*}

\vfill

\end{document}